\documentclass[aps,12pt,prb]{revtex4-1}
\usepackage{graphicx}
\usepackage{subfig}
\usepackage{graphics}
\usepackage{amsfonts}
\usepackage{mathrsfs}
\usepackage{amsmath,dcolumn,epsfig,amssymb,bm,units}
\usepackage{tipa}

\renewcommand \baselinestretch{1.2}

\begin{document}

\begin{titlepage}

\title{
$~~$\\ \vspace{20mm} $~~$\\
A DPF analysis yields accurate analytic potentials for  \\
Li$_2(a\,^3\Sigma^+_u)$ and  Li$_2(1\,^3 \Sigma^+_g)$ 
that incorporate 3-state mixing \\
near the $1\,^3 \Sigma^+_g$-state asymptote}

\author{Nikesh S. Dattani,\footnote{Present
address: Quantum Information Processing Building, Department of Materials, 
12/13 Parks Road, OX1 3PH, UK: dattani.nike@gmail.com} and Robert J. Le
Roy\footnote{leroy@uwaterloo.ca}}

\affiliation{Department of Chemistry, University of Waterloo, Waterloo,
ON~ N2L 3G1, Canada }

\date{\today}

\begin{abstract}
\vspace{10mm}
A combined-isotopologue direct-potential-fit (DPF) analysis of optical
and photoassociation spectroscopy data for the $a\,^{3}\Sigma_u^+$ and
$1\,^3\Sigma_g^+$ states of Li$_2$ has yielded accurate analytic potential
energy functions for both states.  The recommended M3LR$_{5,3}^{8.0}(3)$
potential for the $a\,^3\Sigma_u^+$ of $^{7,7}$Li$_2$ has a well depth of
$\,{\mathfrak D}_e= 333.758(7)\,{\rm cm^{-1}}$ and equilibrium distance
of $\,r_e= 4.17005(3)~$\AA, and the associated scattering lengths are
$\,a_{\rm SL}= -14.759(9)\,$\AA\ for $^{7,7}$Li$_2$ and $-1906(50)\,$\AA\ for
$^{6,6}$Li$_2$.  For the $1\,^3\Sigma_g^+$ state, in spite of a gap of $\,\sim
5200\,{\rm cm}^{-1}$ (from $\,v(1\,^3\Sigma_g^+)= 8-61$) for which there are
no data, the DPF procedure has no difficulty determining an accurate overall
potential. The $1\,^3\Sigma_g^+$ state of the $^{7,7}$Li$_2$ isotopologue
has a well depth of $\,{\mathfrak D}_e= 7092.417(33)\,{\rm cm^{-1}}$ and
equilibrium distance of $\,r_e= 3.06524(9)~$\AA.  The long-range tail of
the recommended M3LR$_{6,3}^{3.6}(9)$ potential energy function for the
$1\,^3\Sigma_g^+$ state is defined by the lowest eigenvalue of a $3\times
3$ long-range interstate coupling matrix to take into account the 3-state
mixing near its asymptote.

\vspace{-10mm}

\end{abstract}
\maketitle

\end{titlepage}

\newpage

\section{Introduction}

Modern theoretical studies of ultra-cold atomic gases demand a very accurate
knowledge of the potential energy curves (PECs) of the systems of interest.
Since Li$_2$ is the second smallest uncharged stable homonuclear molecule,
its chemical and physical properties are particularly interesting.  In recent
years, considerable effort has been focussed on the lowest singlet states of
Li$_2$.\cite{d:dhsu75,d:bara86b,d:urba96,d:lint96,d:mart97,d:wang98a,d:wang02,d:adoh04,d:coxo06a,d:lero09}~
However, the properties of the triplet states of Li$_2$ are much less
well known.

The first observation of discrete spectra involving the lowest triplet state
of Li$_2$ was reported in 1985 by Xie and Field,\cite{d:xxie85} who used
perturbation-facilitated optical-optical double resonance (PFOODR) techniques
to excite $2\,^3\Pi_g -a\,^3\Sigma_u^+$ emission.  They observed transitions
involving $\,v(a\,^3\Sigma_u^+)= 0-6\,$, but because of the limited resolution
available at the time, their results have been superceded by later work.
The first high-resolution triplet-system measurements were reported three
years later by Martin {\em et al.},\cite{d:mart88} who performed a Fourier
transform study of the $1\,^3\Sigma_g^+\to a\,^3\Sigma_u^+$ system of
$^{7,7}$Li$_2$ involving $\,v'=1-7\,$ of the upper state and $\,v''=0-7\,$
of the ground triplet state, with average measurement uncertainties of
only $\pm 0.01~{\rm cm}^{-1}$.  Analogous results for the same system
of $^{6,6}$Li$_2$, spanning the same ranges of vibrational levels, and
with the same accuracy, were reported a year later by Linton {\em et
al.}\cite{d:lint89}~ A decade later Linton {\em et al.}\cite{d:lint99}
reported a high-resolution version of the PFOODR experiment of Xie and
Field\cite{d:xxie85} which yielded accurate (uncertainties ranging
from 0.005 to 0.001 cm$^{-1}$) observations of transitions into
$\,v(a\,^3\Sigma_u^+)\!=\!0-9\,$ from a handful of rotational levels of
the $\,v'=1$ and 2 levels of the $2\,^3\Pi_g$ state of $^{7,7}$Li$_2$.
In addition , a two-photon photoassociation spectroscopy (PAS) experiment
by Abraham {\em et al.}\cite{d:abra95a} had yielded a direct measurement
of the $0.416(\pm 0.001)~{\rm cm}^{-1}$ binding energy of the $\,v\!=\!10$,
$\,N\!=\!0\,$ level of the $a\,^3\Sigma_u^+$ state of $^{7,7}$Li$_2$.

In addition to the seven vibrational levels $\,v=1\!-\!7\,$ of the
$1\,^3\Sigma_g^+$ state observed in the emission experiments, the binding
energies of levels $\,v=62-90\,$ of $^{7,7}$Li$_2$ and $\,v=56-84\,$
of $^{6,6}$Li$_2$ were measured in a PAS study by Abraham {\em et
al.}\cite{d:abra95b}~ However, to date there has been no reported attempt
to bridge the  5100\,cm$^{-1}$ gap between the two sets of results in
order to provide a global description of this state.  This problem is
illustrated by Fig.\,1, which shows the regions of the $a\,^3\Sigma_u^+$
and $1\,^3\Sigma_g^+$ potentials associated with the currently available
experimental data.  The task of bridging the chasm for the $1\,^3\Sigma_g^+$
state is complicated by the fact that in the region very near the upper-state
asymptote spanned by the PAS data, a transition from case\,(b) to case\,(c)
coupling leads to a mixing of the $1\,^3\Sigma_g^+$ state with two other
states that also have $1_g$ symmetry in this long-range ($r\gtrsim 20\,$\AA)
region.

The only PECs which have been reported for the lowest triplet states
($a\,^3\Sigma_u^+$ and $1\,^3\Sigma_g^+$) of Li$_2$, are point-wise,
semiclassical RKR curves generated from Dunham or near-dissociation
expansions for the vibrational energies and inertial rotation $B_v$
constants.\cite{d:mart88,d:lint89,d:lint99}~  For the $1\,^3\Sigma_g^+$
upper state, those PECs were based only on data for vibrational
levels $\,v'=0-7\,$,\cite{d:mart88,d:lint89} since that work
preceded the the photo-association spectroscopy (PAS) studies of this
system.\cite{d:abra95a,d:abra95b}~ Thus, the best available potential
for the $1\,^3\Sigma_g^+$ state provides no realistic predictions for
those subsequently observed high vibrational levels, took no account of
interactions with other states of Li$_2$ near the dissociation asymptote of $1\,^3\Sigma_g^+$,
and did not incorporate the theoretically known inverse-power long-range
behaviour.  Finally, in all studies of these states to date, $^{7,7}\rm{Li}_2$
and $^{6,6}\rm{Li}_2$  were treated independently, and as a consequence, the
effect of Born-Oppenheimer breakdown (BOB) in this system remains unknown.

The present work presents a fully quantum mechanical direct-potential-fit
(DPF) data analysis which accounts for all of the optical and PAS data
described above in terms of global analytic potential energy functions
for the $a\,^3\Sigma_u^+$ and $1\,^3\Sigma_g^+$ states of Li$_2$,
while independent term values are used to represent the 20 levels of the
$2\,^3\Pi_g$ state giving rise to the observed high-resolution emission
into $\,v(a\,^3\Sigma_u^+)= 0-9\,$.  The three longest-range inverse-power
contributions to the interaction energy are incorporated into the
$a\,^3\Sigma_u^+$ and $1\,^3\Sigma_g^+$ state potential energy functions,
and the function for the $1\,^3\Sigma_g^+$ state explicitly accounts for
the three-state mixing referred to above.  In addition, the incorporation
of adiabatic Born-Oppenheimer breakdown (BOB) correction functions in the
Hamiltonians for the $a\,^3\Sigma_u^+$ and $1\,^3\Sigma_g^+$ states allows
the data for the two isotopologues to be treated simultaneously.

Three aspects of this system made its analysis unusually challenging.
Firstly, the long-range tail of the $1\,^3\Sigma_g^+$ state PEC is not the
familiar sum of simple inverse-power terms, since the $1\,^3\Sigma_g^+$
state couples strongly to two other states near the dissociation asymptote
(see \S\,II.C.1).  Secondly, the fact that the leading long-range term of
the PEC tail for the $1\,^3\Sigma_g^+$ state is $C_3/r^3$ means that the
Morse/long-range (MLR) potential function leads to unphysical long-range
behavior if not addressed appropriately (see \S\,II.C.3).\cite{d:lero09}~
Finally, as illustrated by Fig.\,1, the data for the $1\,^3\Sigma_g^+$
state has gaps from $v=7$ to $v=62$ for $^{7,7}\rm{Li}_2$ and from $v=7$ to
$v=56$ for $^{6,6}\rm{Li}_2$, which span more than 72\% of the well depth.
Such a large gap in experimental information has never (to our knowledge)
been treated successfully by a potential-fit analysis in a purely empirical
manner.

\vspace{-4mm}
\section{Models and Methodology}
\vspace{-4mm}
\subsection{DPF Data Analyses and the Radial Hamiltonian }
\vspace{-4mm}

In a DPF spectroscopic data analysis, the upper and lower level of every
observed energy transition is assumed to be an eigenvalue of an effective
radial Schr{\"o}dinger equation characterized by a parameterized potential
energy function and (when appropriate) parameterized radial strength
functions characterizing appropriate BOB terms.  Given some plausible
initial trial parameter values for characterizing the relevant potential,
solution of the associated Schr{\"o}dinger equation yields an eigenvalue
$E_{v,J}$ and eigenfunction $\psi_{v,J}(r)$ for each observed level.
The difference between the energies of appropriate upper and lower levels
then yields an estimate of each observed transition energy, while use of
the Hellmann-Feynman theorem:
\begin{equation}  \label{eq:dEdp}
\frac{\partial E_{v,J}}{\partial p_j} ~=~\left\langle \psi_{v,J}(r) \left|
\frac{\partial \hat H}{\partial p_j} \right| \psi_{v,J}(r)
\right\rangle ,
\end{equation}
yields the partial derivatives required for performing
a least-squares fit of the simulated transitions to the experimental data.

Since the observed transition energies are not linear functions of the
parameters defining the effective radial Hamiltonian, a DPF analysis
requires the use of an iterative non-linear least-squares fitting procedure.
The quality of a given fit is characterized by the value of the dimensionless
root-mean-square deviation of the $N$ experimental data $y_{\rm obs}(i)$
from the predicted values $y_{\rm calc}(i)$ generated from the relevant
Hamiltonian(s):
\begin{equation}  \label{eq:dd}
\overline{dd} ~\equiv~\sqrt{ \frac{1}{N} \sum_{i=1}^N \left( \frac{y_{\rm
calc}(i) -y_{\rm obs}(i)}{u(i)} \right)^2 } ,
\end{equation}
in which $u(i)$ is the uncertainty in the reported value of experimental
datum $i$.  In the present work, these fits were performed using the publicly
available program {\sc DPotFit},\cite{d:DPotFit} while the requisite
initial estimates of the potential function parameters were obtained
using the program \texttt{betaFIT}\cite{d:betaFIT} with preliminary RKR
potentials generated using conventional Dunham expansions.

As with most diatomic DPF analyses reported to date, the present work
is based on the effective radial Schr{\"o}dinger equation presented by
Watson,\cite{d:wats80,d:wats04} and uses the conventions described in
Refs.\,\onlinecite{d:lero99,d:lero02a}.  In particular, the 
rovibrational levels of isotopologue $\alpha$ of diatomic molecule AB in
a given electronic state are the eigenvalues of the radial Schr{\"o}dinger
equation:
\begin{eqnarray}
&&\left\{ -\,\frac{\hbar^2}{2\mu_\alpha}\,\frac{d^2~}{dr^2}\,+\, \left[
  V_{\rm ad}^{(1)}(r)+ \Delta V^{(\alpha)}_{\rm ad} (r) \right]\right.
  \label{eq:schr}\\
&&\left. ~~+~ \frac{\hbar^2\,J(J+1)}{2\mu_\alpha\,r^2} \left[ 1+g^{(\alpha)}(r)
  \right] \right\} \psi_{v,J}(r) ~=~E_{v,J}\, \psi_{v,J}(r) , \nonumber 
\end{eqnarray}
in which $V_{\rm ad}^{(1)}(r)$ is the effective adiabatic internuclear
potential for a selected reference isotopologue labeled $\,\alpha\!=\!1\,$,
$~\Delta V^{(\alpha)}_{\rm ad}(r)=V_{\rm ad}^{(\alpha)}(r) - V_{\rm
ad}^{(1)}(r)~$ is the {\it difference} between the effective adiabatic
potentials for isotopologue--$\alpha$ and isotopologue--1, $g^{(\alpha)}(r)$
is the non-adiabatic centrifugal-potential correction function for
isotopologue--$\alpha$, and the reduced mass $\mu_\alpha$ is defined by the
atomic masses $M_{\rm A}^{(\alpha)}$ and $M_{\rm B}^{(\alpha)}$.  Following
standard conventions,\cite{d:wats80,d:wats04,d:lero99,d:lero02a,d:huan03}
the BOB terms $\Delta V^{(\alpha)}_{\rm ad} (r)$ and $g^{(\alpha)}(r)$
are both written as a sum of contributions from component atoms A and B:
\begin{eqnarray}
\Delta V_{\rm ad}^{(\alpha)} (r) ~&=&~\frac{\Delta M_{\rm
A}^{(\alpha)}} {M_{\rm A}^{(\alpha)}}~\widetilde S_{\rm ad}^{\rm
A}(r)~+~\frac{\Delta M_{\rm B}^{(\alpha)}} {M_{\rm B}^{(\alpha)}}~ 
\widetilde S_{\rm ad}^{\rm B}(r) ,                  \label{eq:dVad}  \\
g^{(\alpha)} (r)~&=&~ \frac{M_{\rm A}^{(1)}} {M_{\rm A}^{(\alpha)}}~ \widetilde
R_{\rm na}^{\rm A}(r)~+~\frac{M_{\rm B}^{(1)}} {M_{\rm B}^{(\alpha)}}~
\widetilde R_{\rm na}^{\rm B}(r) , \label{eq:gnad}
\end{eqnarray}
in which $\,\Delta M_{\rm A/B}^{(\alpha)}= M_{\rm A/B}^{(\alpha)} - M_{\rm
A/B}^{(1)}\,$ are the differences between the atomic masses of atoms A
or B in isotopologue--$\alpha$  and in isotopologue--$1$.  In the present
case \,A\,=\,B\,=\,Li, and these expressions collapse to
\begin{eqnarray}
\Delta V_{\rm ad}^{(\alpha)} (r) ~&=&~\left( \frac{\Delta M_{{\rm
Li}^a}^{(\alpha)}} {M_{{\rm Li}^a}^{(\alpha)}} + \frac{\Delta M_{{\rm
Li}^b}^{(\alpha)}} {M_{{\rm Li}^b}^{(\alpha)}}\right) \widetilde S_{\rm 
ad}^{\rm Li}(r) \label{eq:dVadLi}  \\
g^{(\alpha)} (r)~&=&~ \left( \frac{M_{{\rm Li}^a}^{(1)}} {M_{{\rm
Li}^a}^{(\alpha)}} + \frac{M_{{\rm Li}^b}^{(1)}} {M_{{\rm
Li}^b}^{(\alpha)}}\right) \widetilde R_{\rm na}^{\rm Li}(r) .\label{eq:gnaLi}
\end{eqnarray}
Although only a single radial strength function of each type must be
considered in the present case ($\widetilde S_{\rm ad}^{\rm Li}(r)$ and
$\widetilde R_{\rm na}^{\rm Li}(r)$), two mass factors must be retained in
order to allow us to describe all possible molecular isotopologues.

\vspace{-4mm}
\subsection{The `Basic' Morse/Long-Range (MLR) Potential Energy Function }
\vspace{-4mm}

The next step is to introduce an optimal analytic function for
representing the effective adiabatic internuclear potential for the
reference isotopologue, $\,V_{\rm ad}^{(1)}(r) \equiv V(r)\,$.  The present
work is based on use of the version of the Morse/long-range (MLR)
potential energy function of Refs.\,\onlinecite{d:lero09,d:lero11},
\begin{equation}   \label{eq:VMLR}
V_{\rm MLR}(r) ~=~{\mathfrak D}_e \left[ 1 - \frac{u_{\rm LR}(r)} {u_{\rm
LR}(r_e)}~e^{-\beta(r)\cdot y_p^{\rm{eq}}(r)} \,\right]^2 ~~,
\end{equation}
in which $\mathfrak{D}_e$ is the well depth, $\,r_e\,$ the equilibrium
internuclear distance, and the radial variable in the exponent is
\begin{equation}  \label{eq:ypeq}
y_p^{\rm eq}(r)~\equiv~ \frac{r^p - {r_e}^p} {r^p + {r_e}^p} ~~.
\end{equation}
The parameterized exponent coefficient function $\beta(r)$ which governs
the details of the shape of the potential is defined so that
\begin{equation}   \label{eq:betaINF}
\lim_{r\rightarrow\infty}\beta(r)  ~\equiv~   \beta_\infty 
~=~\ln\left( \frac{2\,{\mathfrak D}_e}{u_{\rm LR}(r_e)}\right) ~~,
\end{equation}
and as a result, the long-range behavior of the potential energy function is
defined by the function $\,u_{\rm LR}(r)\,$:
\begin{equation}  \label{eq:VMLRlim}
V_{\rm MLR}(r) ~\simeq~ {\mathfrak D}_e ~-~ u_{\rm LR}(r) ~+~ \mathcal{O} 
\left(u_{\rm{LR}}(r)^2/4{\mathfrak D}_e \right), \ldots ~~,
\end{equation}
while the denominator factor $\,u_{\rm LR}(r_e)\,$ is simply the value
of that long-range tail function evaluated at the equilibrium bond length.

The theory of long-range intermolecular forces shows us that in general,
$u_{\rm LR}(r)$ may be written in the form
\begin{equation}  \label{eq:uLRdamp}
u_{\rm LR}(r)~=~ \sum_{i=1}^{\rm last}~D_m(r)~\frac{ C_{m_i}}{r^{m_i}} ~~.
\end{equation}
in which the powers $m_i$ and coefficients $C_{m_i}$ of the terms
contributing to this sum are determined by the symmetry of the
given electronic state and the nature of the atoms to which it
dissociates,\cite{d:marg39,d:hirs64,d:hirs67,d:mait81} and the `damping
functions' $D_m(r)$ were introduced to take account of the weakening
of the interaction energies associated with these simple inverse-power
terms due to overlap of the electronic wavefunctions on the interacting
atoms.\cite{d:kree69}~ While most previous applications of the MLR
potential function form omitted the $D_m(r)$ damping function factors, it
was shown in Ref.\,\onlinecite{,d:lero11} that in addition to providing
a more realistic physical description of the long-range potential tail,
their introduction improves the extrapolation behaviour of the repulsive
short-range potential wall, and when they are included, fewer parameters
are required to achieve a given quality of fit to experimental data.
In either case, the structure of Eq.(\ref{eq:VMLRlim}) means that at large
distances where $\,u_{\rm{LR}}(r) \gg u_{\rm LR}(r)^2/(4\,{\mathfrak D}_e)$,
the long-range behaviour of $V_{\rm{MLR}}$ is defined by $u_{\rm{LR}}(r)$.

Following Ref.\,\onlinecite{d:lero11}, the present work uses the modified
Douketis-type\cite{d:douk82} damping function form
\begin{equation}  \label{eq:DSdamp}
D_m^{\textrm{DS}(s)}(r) ~=~ \left(\, 1 ~-~ e^{-\,\frac{b^{\rm ds}(s)\,(\rho\,r)}{m}\,
-\, \frac{c^{\rm ds}(s)\cdot(\rho\,r)^2}{\sqrt{m}} }\,\right)^{m + 1} ~~,
\end{equation}
with $s=-1$. Here, $\,b^{\rm ds}(s)$ and $\,c^{\rm ds}(s)$ are treated as
system-independent parameters with $\,b^{\rm ds}(-1)= 3.30\,$ and $\,c^{\rm
ds}(-1)=0.423\,$. For interacting atoms A and B, $\rho\equiv \rho_{\rm AB} =
2\rho_{\rm A}\rho_{\rm B}/(\rho_{\rm A} + \rho_{\rm B})$, in which $\rho_{\rm
A} = (I^{\rm A}_p/I^{\rm H}_p)^{2/3}$ is defined in terms the ionization
potential of the atom A and that of an H atom ($I^{\rm A}_p$ and $I^{\rm
H}_p$ respectively).  Inclusion of these damping functions means that at
very short distances $V_{\rm MLR}(r) \propto 1/r^2\,$.\cite{d:lero11}~
Comparisons with {\em ab initio} results for a sampling of chemical and Van
der Waals interactions showed that this type of damping function yielded
quite realistic MLR short-range extrapolation behaviour,\cite{d:lero11}
so this $D_m(r)$ form is adopted here.

In order to ensure that the exponent coefficient function in
Eq.\,(\ref{eq:VMLR}) satisfies Eq.\,(\ref{eq:betaINF}), it is customary
to write it as the constrained polynomial:
\begin{equation}   \label{eq:betapq}
\beta(r)~=~ \beta_p^q(r)~\equiv~ y_p^{\rm ref}(r)\,\beta_\infty ~+~ \left[1 -
y_p^{\rm ref}(r)\right] ~\sum_{i=0}^{N} \beta_i \,[y_q^{\rm ref}(r)]^i ~~.
\end{equation}
This function is expressed in terms of two radial variables which are similar
to $y_p^{\rm eq}(r)$, but are defined with respect to a different expansion
center ($r_{\rm ref}$), and involve two different powers, $p$ and $q$ (the
reasons for this structure are discussed in Ref.\,\onlinecite{d:lero09}):
\begin{equation}  \label{eq:ypq_ref}
y_p^{\rm ref}(r)~\equiv~ \frac{r^p - {r_{\rm ref}}^p}
{r^p + {r_{\rm ref}}^p} \hspace{20mm} {\rm and}  \hspace{20mm}
y_q^{\rm ref}(r)~\equiv~ \frac{r^q - {r_{\rm ref}}^q}
{r^q + {r_{\rm ref}}^q}  ~~.
\end{equation}
The limiting long-range behaviour of the exponential term in
Eq.\,(\ref{eq:VMLR}) gives rise to additional inverse-power contributions
to the long-range potential of Eq.\,(\ref{eq:VMLRlim}), with the leading
term being proportional to $1/r^{m_1+p}$.  This means that the power $p$
must be greater than $\,(m_{\rm last} - m_1)\,$ if the long-range behaviour
of Eq.\,(\ref{eq:uLRdamp}) is to be maintained.\cite{d:lero09}~
There is no analogous formal constraint on the value of $\,q\,$; however,
experience suggests that its value should lie in the range $\,2~\lesssim
~q~\leq p\,$.\cite{d:lero09,d:coxo10,d:lero11}~  In early work with this
potential function form, the radial variables in Eq.\,(\ref{eq:betapq}) were
both defined as $y_p^{\rm eq}(r)$ of Eq.\,(\ref{eq:ypeq}) (i.e., $\,r_{\rm
ref}=r_e\,$ and $\,q=p$).\cite{d:lero06c,d:lero07,d:sala07,d:fxie09}~
However, it has since been shown that setting $\,r_{\rm{ref}}
> r_e\,$ and $\,q<p\,$ can significantly reduce the number of $\beta_i$ parameters required
to describe a given data set accurately, and yields more stable
expansions.\cite{d:lero09,d:coxo10,d:lero11}~

A second consideration associated with the use of the damping functions
of Eq.\,(\ref{eq:DSdamp}) is their effect on the shape of the short-range
repulsive potential wall of an MLR potential.  As was pointed out in
Ref.\,\onlinecite{d:lero11}, the fact that the radial variables $y_{p/q}^{\rm
ref}(r) \,\to -1\,$ as $\,r\to 0\,$ means that at very small distances
$\,V_{\rm MLR}(r) \propto \{u_{\rm LR}(r)\}^2$.  If damping functions are
neglected (i.e., if Eq.\,(\ref{eq:uLRdamp}) did not include the $D_m(r)$
functions), then the limiting short-range behaviour of the potential
energy function would be $\,V_{\rm MLR}(r)\propto 1/r^{2m_{\rm last}}$.
However, for a typical two- or three-term $u_{\rm LR}(r)$ expansion,
$\,m_{\rm last}= 8\,$ or 10, and the resulting $1/r^{16}$ or $1/r^{20}$
short-range repulsive wall behaviour would be unphysically excessively steep.
In the data-sensitive region of the potential well, this excessive growth
rate would be compensated for by the behaviour of the empirically determined
exponent coefficient function $\beta(r)$.  However, the unphysical high-order
$\,r^{-16}$ or $r^{-20}\,$ singular behaviour would re-assert itself in
the shorter-range extrapolation region.

In this paper, the label for particular MLR potential function models is
written as M$x$LR$_{p,q}^{r_{\rm{ref}}}(N)$, in which $x$ is the number
of inverse-power terms incorporated into $u_{\rm{LR}}(r)$, while $p$,
$q$, $r_{\rm{ref}}$ and $N$ are defined above.  The `basic' MLR model
described above is used herein to describe the potential energy function
for the $a\,^3\Sigma_u^+$ state of Li$_2$.  However, some enhancements
were required for treating the $1\,^3\Sigma_g^+$ state.

\vspace{-4mm}
\subsection{Modified MLR Potential for the $1\,^3\Sigma_g^+$ State of
Li$_2$}
\vspace{-4mm}
\subsubsection{Incorporating interstate coupling into the MLR model }
\label{coupling}
\vspace{-4mm}

At a preliminary stage of the present work, the $1\,^3\Sigma_g^+$
state of Li$_2$ was represented by the `basic' MLR function of
Eqs.\,(\ref{eq:VMLR})--(\ref{eq:ypq_ref}), in which $u_{\rm LR}(r)$
consisted of three terms, with $\,m_i=\{3, 6, 8\}$.  Because of the added
complexity due to interstate coupling near the dissociation asymptote,
all damping functions for this state were fixed at $D_m(r)=1$. This model
was able to provide an excellent fit both to the fluorescence data for
$\,v'=0-7$, and to the $^{7,7}$Li$_2$ PAS data for $\,v=62-70$ and the
$^{6,6}$Li$_2$ PAS data for $\,v=56-65$ whose upper limits which correspond
to binding energies of about 24~cm$^{-1}$.  However, when PAS data for
higher vibrational levels were included in the analysis, the quality of
fit got progressively worse, and the discrepancies could not easily be
removed simply by increasing the order of the the polynomial $\beta(r)$.
The reason for this increasing inability of the basic MLR model to account
for levels lying very near dissociation is that the $1\,^3\Sigma_g^+$
state of Li$_2$ couples to two other states near its dissociation asymptote.

This same type of problem was encountered in a recent study
of the $A(^1\Sigma_u^+)-X(^1\Sigma_g^+)$ system of Li$_2$.
In that case the $0_u^+(A\,^1\Sigma_u^+)$ state which goes to
the Li$(^2P_{\nicefrac{1}{2}})+{\rm Li}(^2S_{\nicefrac{1}{2}})$
asymptote couples to the $0_u^+(b\,^3\Pi)$ state which goes to the
higher Li$(^2P_{\nicefrac{3}{2}})+{\rm Li}(^2S_{\nicefrac{1}{2}})$
limit,\cite{d:movr77,d:aube98a} and the energies of levels lying near
dissociation could not be explained properly without taking account of
the inter-state mixing.  Fortunately, Aubert-Fr{\' e}con and co-workers
had derived an analytic description of those coupled states based on the
eigenvalues of a $2\!\times\!2$ interaction matrix,\cite{d:mart97,d:aube98a}
and it was shown that their analytic expression for the lower eigenvalue
could readily be used to define $u_{\rm LR}(r)$ for this state in an MLR
potential model.\cite{d:lero09}~

Treatment of levels lying near the dissociation limit of the
$1\,^3\Sigma_g^+$ state of Li$_2$ involves a similar problem;
while it dissociates to the Li$(^2P_{\nicefrac{1}{2}})+{\rm
Li}(^2S_{\nicefrac{1}{2}})$ limit, it couples to $1_g(^1\Pi)$
and $1_g(^3\Pi)$ states which correlate with the higher
Li$(^2P_{\nicefrac{3}{2}})+{\rm Li}(^2S_{\nicefrac{1}{2}})$
limit.\cite{d:movr77,d:aube98a}~ Since the Li($^2P$) state spin-orbit
splitting is quite small (ca.\ $0.335~{\rm cm}^{-1}$), the interstate
coupling only becomes important for levels lying relatively close to
the dissociation limit.  Fortunately, Aubert-Fr{\'e}con and co-workers
have studied this case too.\cite{d:aube98a}~ In particular, they presented
expressions for the six independent elements of the symmetric $3\!\times\!3$
matrix that defines the long-range interaction energies for these
three states.  Their matrix elements took into account the first-order
resonance-dipole ($1/r^3$) term, the leading dispersion energy terms,
and the exchange energy.  If we neglect the exchange term, keep only the
first two ($m=6$ and 8) dispersion energy terms, set the zero of energy at
the $1\,^3\Sigma_g^+$--state asymptote, make use of the symmetry relation
for $\,m=3\,$,
\begin{equation}   \label{eq:C3sigma}
C_3^{^3\Sigma_g^+}~=~ 2\,C_3^{^1\Pi_g} ~=~-2\,C_3^{^3\Pi_g}~\equiv ~C_3^\Sigma
~~,
\end{equation}
and that for $\,m=6\,$,
\begin{equation}    \label{eq:CnPi}
C_6^{^1\Pi_g}~=~C_6^{^3\Pi_g}~\equiv~C_6^\Pi ~~,
\end{equation}
and define $\,C_{6,8}^\Sigma \equiv C_{6,8}^{^3\Sigma_g^+}$, their
$3\!\times\!3$ long-range interaction matrix ${\bf M}_{\rm LR}$ becomes
{\footnotesize
\begin{equation}  \label{eq:matrix} \hspace{-10mm}
\left( \begin {array}{ccc} -\frac{1}{3} \left(\frac{C_3^\Sigma}
  {r^3}+\frac{C_6^\Sigma+ 2 C_6^\Pi}{r^6}+ \frac{C_8^\Sigma+ C_8^{^1\Pi_g} +
  C_8^{^3\Pi_g}}{r^8} 
  \right) &
\frac{\sqrt{2}}{3} \left(\frac {C_3^\Sigma}{r^3}+ \frac{C_6^\Sigma- C_6^\Pi}
  {r^6}+\frac{2C_8^\Sigma- C_8^{^1\Pi_g} - C_8^{^3\Pi_g}}{2r^8} \right) &
\frac{1}{\sqrt{6}} \left( \frac{C_3^\Sigma}{r^3} + \frac{C_8^{^1\Pi_g} -
  C_8^{^3\Pi_g}}{r^8} \right) \\
\noalign{\medskip} 
\frac{\sqrt{2}}{3} \left(\frac{C_3^\Sigma}{r^3}+ \frac{C_6^\Sigma- C_6^\Pi}
  {r^6}+\frac{2C_8^\Sigma- C_8^{^1\Pi_g} - C_8^{^3\Pi_g}}{2r^8} \right) &
-\frac{2}{3} \left(\frac{C_3^\Sigma}{r^3} + \frac{2C_6^\Sigma + C_6^\Pi}{2r^6}
  + \frac{4C_8^\Sigma + C_8^{^1\Pi_g} + C_8^{^3\Pi_g}}{4r^8} \right) 
  + \Delta E &
\frac{1}{2\sqrt{3}} \left( \frac{C_3^\Sigma}{r^3} + \frac{C_8^{^1\Pi_g} -
  C_8^{^3\Pi_g}}{r^8} \right) \\
\noalign{\medskip}
\frac{1}{\sqrt{6}} \left( \frac{C_3^\Sigma}{r^3} + \frac{C_8^{^1\Pi_g} -
  C_8^{^3\Pi_g}}{r^8} \right) &
\frac{1}{2\sqrt{3}}  \left( \frac{C_3^\Sigma}{r^3} + \frac{C_8^{^1\Pi_g} -
  C_8^{^3\Pi_g}}{r^8} \right) &
-\left(\frac{C_6^\Pi}{r^6} + \frac{C_8^{^1\Pi_g} + C_8^{^3\Pi_g}}{2r^8} \right)
  + \Delta E 
\end{array}
\right)~~,
\end{equation}  }
in which $\Delta E$ is the accurately known (positive) spin-orbit
splitting energy of Li($^2P$).\cite{d:sans95}~ Note that in contrast with
Ref.\,\onlinecite{d:aube98a}, the present formulation treats attractive $C_m$
coefficients as positive, rather than negative quantities.  Following the
correlation scheme given by Movre and Pichler,\cite{d:movr77} the lowest
eigenvalue of the matrix \eqref{eq:matrix} defines the long-range tail of
the $1\,^3\Sigma_g^+$-state interaction potential; see Fig.\,2.

Analytic expressions for the three eigenvalues of Eq.\,(\ref{eq:matrix}) were
reported in Ref.\,\onlinecite{d:aube98a}.  Those expressions for the zeros
of the characteristic polynomial for ${\bf M}_{\rm LR}$ were obtained using
the method of Scipione del Ferro and Niccol\`{o} Fontana Tartaglia (first
published\cite{d:card1545} by Gerolamo Cardano in 1545), and by applying a
trigonometric substitution to avoid expressions involving square-roots of
negative quantities.  However, Kopp has demonstrated that while it is useful
for obtaining symbolic expressions, this formula can yield substantial
errors when used for actual computations, primarily because of the
numerical errors that accumulate when computing the arctan function within
the formula.\cite{d:kopp08}~ Moreover, the symbolic expressions for the
derivatives of the lowest eigenvalue with respect to the $C_n^{\Sigma/\Pi}$
coefficients required by the least-squares fitting procedure are
inconveniently complex.  Because of these problems, in the present work
the eigenvalues of this interaction matrix were calculated numerically
(using the Jacobi eigenvalue algorithm\cite{d:jaco1846,d:kopp08}), and
their derivatives with respect to the $C_n^{\Sigma/\Pi}$ coefficients were
computed using the discrete version of the Hellmann-Feynman theorem:
\begin{equation}   \label{eq:HFthm}
\frac{d \lambda_i}{d p} = \left\langle \phi_\lambda \left|  \frac{d {\bf
M}_{\rm LR}}{dp} \right| \phi_\lambda \right\rangle  ,
\end{equation} 
in which $\lambda_i$ is the appropriate eigenvalue of the matrix ${\bf
M}_{\rm LR}$, and $\phi_{\lambda_i}$ is the corresponding eigenvector.

\vspace{-4mm}
\subsubsection{Simplifying the treatment of interstate coupling for
Li$_2(1\,^3\Sigma_g^+)$  }
\label{simplifying}
\vspace{-4mm}

The treatment of the long-range behaviour of the $A\,^1\Sigma_u^+$ state
of Li$_2$ in Ref.\,\onlinecite{d:lero09} was precisely analogous to that
for the $1\,^3\Sigma_g^+$ state discussed here, except that while that
case involved a $2\!\times\!2$ matrix whose eigenvalues were determined
analytically, the present case involves the $3\!\times\!3$ matrix of
Eq.\,(\ref{eq:matrix}) whose eigenvalues are calculated numerically.  In the
treatment of the $A\,^1\Sigma_u^+$ state, it was shown that the $C_6^\Pi/r^6$
and $C_8^\Pi/r^8$ terms had virtually no effect on the lower ($\Sigma$-state)
eigenvalue of the $2\!\times\!2$ long-range interstate coupling matrix.
This led us to consider making the same simplification here.

Following the approach of Ref.\,(\onlinecite{d:lero09}), we compared
the values of the lowest eigenvalue of Eq.\,(\ref{eq:matrix}) obtained
when all $C_6^{\Sigma/\Pi}$ and $C_8^{\Sigma/\Pi}$ coefficients were
defined by the theoretical values of Tang {\em et al.},\cite{d:tang09}
with those obtained on setting $\,C_6^\Pi= C_8^{^1\Pi_g}= C_8^{^3\Pi_g}=
0\,$.  Over the range $\,r=2\,$ to 500\,\AA, the difference between these
two estimates of the lowest eigenvalue were always less than $3\times
10^{-6}\,{\rm cm}^{-1}$.  Thus, it seems clear that in the present treatment
of the $1\,^3\Sigma_g^+$ state of Li$_2$, no significant errors will
be introduced if contributions involving $C_6^\Pi$, $C_8^{^1\Pi_g}$ and
$C_8^{^3\Pi_g}$ are omitted from Eq.\,(\ref{eq:matrix}).  At the same time,
it is important to note that these $C_6^\Pi$ and $C_8^\Pi$ coefficients
{\em cannot}\, be neglected when using the two higher eigenvalues of
Eq.\,(\ref{eq:matrix}) to define the long-range tails of the $^3\Pi_g$
and $^1\Pi_g$ states which couple with the $1\,^3\Sigma_g^+$ state of
interest here.  This point is illustrated by Fig.\,3, which compares plots of
the three eigenvalues of Eq.\,(\ref{eq:matrix}) obtained using all of the
$C_n^{\Sigma/\Pi}$ coefficients of Tang {\em et al.}\cite{d:tang09} (solid
red curves) with those obtained from this same matrix when $\,C_6^\Pi= 0=
C_8^{^1\Pi_g}= C_8^{^3\Pi_g}\,$ (dashed blue curves).  It is clear that at
the smaller distances where the $C_6$ and $C_8$ terms become important,
one cannot use the above approximation when calculating the eigenvalues
of \eqref{eq:matrix} associated with the two $\Pi_g$ states.

\vspace{-4mm}
\subsubsection{Implications of the quadratic term in the MLR potential
function form }
\label{quadratic}
\vspace{-4mm}

It was shown in Ref.\,\onlinecite{d:lero09} that contributions from the
quadratic term in Eq.\,(\ref{eq:VMLRlim}) can give rise to spurious
perturbations in long-range behavior of the MLR potential function form.  In
the present case, the leading terms in the long-range potential for
the $1\,^3\Sigma_g^+$ state of Li$_2$ correspond to $\,m_i=\{3, 6, 8\}$.
If we temporarily ignore the effects of damping and interstate coupling in
order to write $u_{\rm LR}(r)$ as a simple inverse-power sum, the presence
of the quadratic term in Eqs.\,(\ref{eq:VMLR}) and (\ref{eq:VMLRlim})
mean that the effective long-range behavior of the MLR potential would be
\begin{equation}  \label{eq:VLRexp}
V_{\rm MLR}(r) ~\simeq~{\mathfrak D}_e~-~\frac{C_3}{r^3}~-~\frac{C_6}{r^6}~-~
\frac{C_8}{r^8} ~+~\frac{(C_3)^2/(4{\mathfrak D}_e)}{r^6}~+~
\frac{C_{3\,}C_6/(2{\mathfrak D}_e)}{r^9} ~+~ \ldots ~~.
\end{equation}
Thus, if the overall effective long-range behavior is to be defined
by an inverse-power sum governed by the specified $C_3$, $C_6$
and $C_8$ coefficients (and {\em not} include the last two terms in
Eq.\,(\ref{eq:VLRexp})!), the definition of $u_{\rm LR}(r)$ must compensate
for the quadratic terms by being defined as
\begin{equation}  \label{eq:uLRcor}
u_{\rm{LR}}(r)~=~\frac{C_3}{r^3}~+~\frac{C_6^{\rm adj}}{r^6}~+~
\frac{C_{8}}{r^{8}} ~+~ \frac{C_9^{\rm adj}}{r^9}~~,
\end{equation}
in which $\,C_6^{\rm adj} \equiv C_6+(C_3)^2/(4 \mathfrak{D}_e)\,$ and
$\,C_9^{\rm adj}\equiv C_3\,C_6^{\rm adj}/(2 \mathfrak{D}_e)$.

Since the long-range tail of our $1\,^3\Sigma_g^+$-state potential also
includes interstate coupling, these expressions for $C_6^{\rm{adj}}$ and
$C_9^{\rm{adj}}$ (which were derived analytically for potentials with simpler
long-range tails) need to be tested in order to determine how well they
cancel the effect of the spurious last-two terms in Eq.\,(\ref{eq:VLRexp}).
Results of a numerical test of this question are presented in Fig.\,4, which
displays plots of the quantity $\,C_3^{\rm eff}(r)\equiv r^3[{\mathfrak
D}-u_{\rm LR}(r)]\,$ {\em vs.}\ $r^{-3}$ for three different definitions of
$u_{\rm LR}(r)$.  This type of plot illustrates the nature of the long-range
interaction on a reduced scale.  If $u_{\rm LR}(r)$ was defined as the
simple inverse-power sum of Eq.\,(\ref{eq:uLRcor}), as $\,r^{-3}\to 0\,$
the resulting plot would approach an intercept of $C_3$ with a limiting
slope of $C_6^{\rm eff}$.  In the present case, however, the three-state
coupling near the potential asymptote causes all of the plots in Fig.\,4
to drop off sharply for $\,r^{-3}\lesssim 10^{-4}\,{\rm \AA}^{-3}$.

The desired long-range behavior is achieved when $u_{\rm LR}(r)$ is defined
simply as the lowest eigenvalues of the matrix of Eq.\,(\ref{eq:matrix}),
$\,u_{\rm LR}(r)= - \lambda_{\rm min}(r)$.  For the case in which $\,C_6^\Pi
= C_8^{^1\Pi_g}= C_8^{^3\Pi_g}=0\,$ and the other $C_n^{\Sigma/\Pi}$
coefficients are fixed at the values of Tang {\em et al.},\cite{d:tang09}
this desired behavior is defined by the solid black curve in Fig.\,4.
The dash-dot-dot red curve in Fig.\,4 then shows how the long-range behavior
of the associated MLR potential, which includes the quadratic term of
Eq.\,(\ref{eq:VMLRlim}), deviates from this desired long-range behavior.
Next, the dotted blue curve shows the effect on the long-range MLR behavior
of replacing $C_6^\Sigma$ by the quantity $C_6^{\rm adj}$ defined above.
It is immediately clear that this removes most of the discrepancy with
the `ideal' long-range behavior (solid black curve).  Finally, the dashed
green curve shows the effect on the long-range MLR potential tail of also
including the $C_9^{\rm adj}{/r^9}$ term in the definition of $u_{\rm LR}(r)$
in order to cancel out the spurious $r^{-9}$ term in Eq.\,(\ref{eq:uLRcor}).
It is clear that use of the resulting definition
\begin{equation} \label{eq:uLRadj}
u_{\rm LR}(r) ~=~ -\,\lambda_{\rm min}(C_3,C_6^{\rm adj},C_8;r)~+~C_9^{\rm
adj} /r^9
\end{equation}
brings the long-range tail of the overall MLR potential function into
essentially exact agreement with the desired form.

\vspace{-4mm}
\subsubsection{Inclusion of retardation in the model potential for 
Li$_2(1\,^3\Sigma_g^+)$  }
\vspace{-4mm}

It has long been known that at the very large distances where the $C_3/r^3$
term comes to dominate the interaction energy in this type of system,
``retardation'' effects due to the finite speed of light should not be
neglected.\cite{d:mclo64,d:meat68}~ It was shown by Meath\cite{d:meat68}
that the effect of retardation on an $s/p$ resonance-dipole interaction
can be accounted for by multiplying $C_3^\Sigma$ by the function $f_{\rm
ret}^\Sigma(r)$ and $C_3^\Pi$ by the function $f_{\rm ret}^\Pi(r)$, where
\begin{eqnarray}
f_{\rm ret}^\Sigma &=& \cos \left(\frac{r}{\textrm{\textcrlambda}_{SP}}\right)
   + \left(\frac{r}{\textrm{\textcrlambda}_{SP}}\right) \sin\left(\frac{r}
   {\textrm{\textcrlambda}_{SP}} \right)~~ \label{eq:retS}  \\
f_{\rm ret}^\Pi &=& f_{\rm ret}^\Sigma ~-~ \left(\frac{r}
  {\textrm{\textcrlambda}_{SP}}\right)^2 \cos\left(\frac{r}
  {\textrm{\textcrlambda}_{SP}}\right)~~, \label{eq:retP}
\end{eqnarray}
in which ${\textrm{\textcrlambda}_{SP}}=\lambda_{SP}/2\pi\,$ and
$\lambda_{SP}$ is the wavelength of light associated with the atomic
$^2S-{^2P}$ transition.  

It is a straightforward matter to incorporate this retardation behavior
into the MLR potential function form.  In particular, on setting
$\,C_6^{\Pi}=C_8^{\Pi}=0$, making use of the symmetry relationships
among the $C_3^\Sigma$, $C_3^{^1\Pi_g}$ and $C_3^{^3\Pi_g}$ coefficients,
and replacing $C_6^{\Sigma}$ by $C_6^{\Sigma,\rm{adj}}$, the long-range
interstate coupling matrix for the three $1_g$ states dissociating to
yield Li($^2S_{\nicefrac{1}{2}}) + {\rm Li}(^2P)\,$ becomes
\begin{equation}  \label{eq:matrix2}
{\bf M}_{\rm LR} ~=~ \left(\begin{array}{ccc} 
-\frac{1}{3} \left(\frac {C_3^\Sigma \,f_{\rm ret}^\Sigma}{r^3} + 
  \frac{C_6^{\Sigma,\rm{adj}}}{r^6} + \frac {C_8^{\Sigma}}{r^8} \right) & 
\frac{\sqrt{2}}{3} \left(\frac{C_3^\Sigma\,f_{\rm ret}^\Sigma}{r^3} + 
  \frac{C_6^{\Sigma,\rm{adj}}}{r^6} + \frac{C_8^{\Sigma}}{{r}^{8}} \right) & 
\frac{1}{\sqrt{6}}~{\frac{C_3^\Sigma\,f_{\rm ret}^\Pi}{r^3}}  \\
\noalign{\medskip} 
\frac{\sqrt{2}}{3} \left(\frac{C_3^\Sigma\,f_{\rm ret}^\Sigma}{r^3} + 
  \frac{C_6^{\Sigma,\rm{adj}}}{r^6} + \frac{C_8^{\Sigma}}{r^8} \right) & 
-\frac{2}{3} \left(\frac{C_3^\Sigma\,f_{\rm ret}^\Sigma}{r^3} + 
  \frac{C_6^{\Sigma,\rm{adj}}}{r^6}+ \frac{C_8^\Sigma}{r^8} \right)+ \Delta E~&
\frac{1}{2\sqrt{3}}~\frac{C_3^\Sigma\,f_{\rm ret}^\Pi}{r^3}   \\
\noalign{\medskip}
\frac{1}{\sqrt{6}}~\frac{C_3^\Sigma\,f_{\rm ret}^\Pi}{r^3} &
  \frac{1}{2\sqrt{3}}\,\frac{C_3^\Sigma\,f_{\rm ret}^\Pi}{r^3} &
  \Delta E      \end {array} \right)~~.
\end{equation}
Unless stated otherwise, the definition of the long-range tail of the
MLR potential for the $1\,^3\Sigma_g^+$ used throughout the rest of
this study is therefore given by
\begin{equation}  \label{eq:uLRfinal}
u_{\rm LR}^{\rm ret}(r) ~=~ -\,\lambda_{\rm min}^{\rm ret}(C_3^\Sigma,C_6^{\rm
adj},C_8^\Sigma;r) ~+~ C_9^{\rm adj} /r^9
\end{equation}
in which $\lambda_{\rm min}^{\rm ret}(C_3^\Sigma,C_6^{\rm adj},C_8^\Sigma;r)$
is the lowest eigenvalue of the interaction energy matrix of
Eq.\,(\ref{eq:matrix2}).

\vspace{-4mm}
\subsection{Born-Oppenheimer Breakdown Functions}
\vspace{-4mm}

The radial strength functions in Eqs.\,(\ref{eq:dVadLi}) and
(\ref{eq:gnaLi}) may be written as polynomials constrained to have specified
asymptotic values using the format of Eq.\,(\ref{eq:betapq})
\begin{eqnarray}
\widetilde S_{\rm ad}^{\rm Li}(r) &=&y_{p_{\rm ad}}^{\rm eq}(r)~u_\infty^{\rm
  Li} ~+~[1 - y_{p_{\rm ad}}^{\rm eq}(r)]~\sum_{i=0}\, u_i^{\rm Li}~ y_{q_{\rm
  ad}}^{\rm eq}(r)^i   \label{eq:Sad} \\
\widetilde R_{\rm na}^{\rm Li}(r) &=&y_{p_{\rm na}}^{\rm eq}(r)~t_\infty^{\rm
  Li} ~+~ [1 - y_{p_{\rm na}}^{\rm eq}(r)]~\sum_{i=0}\, t_i^{\rm Li}~
  y_{q_{\rm na}}^{\rm eq}(r)^i \label{eq:Rna}
\end{eqnarray}
in which $u_\infty^{\rm Li}$ and $t_\infty^{\rm Li}$ are the values
of these functions in the limit $\,r\to \infty\,$, $u_0^{\rm Li}$
and $t_0^{\rm Li}$ define their values at $\,r=r_e\,$, and the radial
variables are versions of Eq.\,(\ref{eq:ypeq}) associated with chosen
values of the integers $p_{\rm ad}$, $p_{\rm na}$, $q_{\rm ad}$ and $q_{\rm
na}$.\cite{d:lero02a}~  The discussion of Ref.\,\onlinecite{d:lero02a} shows
that $\,t_\infty^{\rm A}=0.0\,$ for any molecule which dissociates to yield an
uncharged atom-A, so $\,t_\infty^{\rm Li}(a\,^3\Sigma_u^+)= t_\infty^{\rm
Li}(1\,^3\Sigma_g^+)= 0.0$.  In addition, we adopt the Watson convention of
setting the parameter $\,t_0^{\rm Li}=0.0\,$ for both the $a\,^3\Sigma_u^+$
and $1\,^3\Sigma_g^+$ states, since its value cannot be determined from
transition-frequency data alone.\cite{d:wats80,d:lero02a,d:wats04}~

Following standard conventions,\cite{d:lero99} the absolute zero of energy is
defined as the energy of ground-state atoms separated at $\,r\to \infty\,$,
so by definition $\,u_\infty^{\rm Li}(a\,^3\Sigma_u^+)= 0.0$.  Since the
$1\,^3\Sigma_g^+$ state of Li$_2$ dissociates to one ground-state
($^2S_{1/2}$) and one excited-state ($^2P_{1/2}$) atom, the value of
$\,u_\infty^{\rm Li}(1\,^3\Sigma_g^+)$ is then defined by the difference
between the atomic $\,^2P_{1/2}\leftarrow {^2S}_{1/2}\,$ excitation energies
for $^6$Li and $^7$Li, which is\cite{d:sans95,d:lero99}
\begin{equation}
\delta E^{\rm ^6Li}_{\rm ^7Li}({^2P}_{1/2})~=~ \Delta E^{\rm ^6Li}(
{^2P}_{1/2}\leftarrow {^2S}_{1/2})~-~\Delta E^{\rm ^7Li} (^2P_{1/2}\leftarrow
{^2S}_{1/2}) ~=~-0.351\,338~[{\rm cm}^{-1}]     \label{eq:dElim}
\end{equation}
This is the difference between the energy asymptotes of the
$1\,^3\Sigma_g^+$ states of $^{6,6}$Li$_2$ and $^{7,7}$Li$_2$,
and it defines the asymptotic value of the adiabatic radial strength
function.\cite{d:lero99}~  Since we select $^{7,7}$Li$_2$ as the reference
isotopologue, this yields
\begin{equation}
u_\infty^{\rm Li}(1\,^3\Sigma_g^+) ~=~ \delta E^{\rm ^6Li}_{\rm ^7Li}
({^2P}_{1/2})\left/ 2\left( 1 - \frac{M({\rm ^7Li})}{M({\rm ^6Li})}
\right) \right.  =~ 1.05574~[{\rm cm}^{-1}]   \label{eq:uainf}
\end{equation}

We now address the choice of powers $p_{\rm ad}$, $q_{\rm ad}$, $p_{\rm
na}$ and $q_{\rm na}$ for defining the radial expansion variables
in Eqs.\,(\ref{eq:Sad}) and (\ref{eq:Rna}).  As was pointed out in
Ref.\,\onlinecite{d:lero99}, if the effective adiabatic potential for the
`minor' isotopologues is to have the same limiting long-range behavior as
that for the reference isotopologue, $p_{\rm ad}$ must be greater than
or equal to the power of the longest-range term in the intermolecular
potential for that state.  Thus, we set $\,p_{\rm ad} (a\,^3\Sigma_g^+)
=6\,$ and $\,p_{\rm ad} (1\,^3\Sigma_g^+) =3$.  Note that BOB radial strength
functions are relatively weak and slowly varying, and few terms are required
to define them.  As a result, there is no need here to introduce an $\,r_{\rm
ref} \neq r_e\,$ extension into the definition of the expansion variables
in Eqs.\,(\ref{eq:Sad}) and (\ref{eq:Rna}), and for the sake of simplicity
we set $\,q_{\rm ad} (a\,^3\Sigma_u^+) = p_{\rm ad} (a\,^3\Sigma_u^+) =6\,$
and $\,q_{\rm ad} (1\,^3\Sigma_g^+) = p_{\rm ad}(1\,^3\Sigma_g^+) =3\,$.

We are not aware of any theoretical predictions regarding the limiting
long-range behavior of the centrifugal non-adiabatic radial strength
function $\widetilde R_{\rm ad}^A(r)$, so we have no basis for assigning
particular values to $p_{\rm na}$.  Moreover, as in the discussion of
\S\,III.B.(iii), there are no physical constraints on the values of
$q_{\rm na}$.  At the same time, Fig.\,3 of Ref.\,\onlinecite{d:lero02a}
shows that use of too small values for these powers can give rise to
physically implausible extrema in the resulting functions on the interval
between the data region and the asymptote, while use of too high values will
lead to a requirement for an excessive number of expansion coefficients.
For simplicity, we therefore chose to set $\,p_{\rm na} = q_{\rm na}=
3\,$ in fits to models which included non-zero $\widetilde R_{\rm na}^{\rm
Li}(r)$ functions for either electronic state.

Since $\,y_{p_{\rm ad}}^{\rm ref}(r_e) =0\,$ and $\,y_{p_{\rm ad}}^{\rm
ref}(r\to \infty)=1$, the algebraic form of Eq.\,(\ref{eq:Sad}) means that
the difference between the well depths of different Li$_2$ isotopologues
in a given electronic state is given by the expression\cite{d:lero09}
\begin{equation}  \label{eq:dDe}  
\delta{\mathfrak D}_e^{(\alpha)}~=~ {\mathfrak D}_e^{(\alpha)} - {\mathfrak
D}_e^{(1)} ~=~\left ( \frac{\Delta M_{{\rm Li}^a}^{(\alpha)}}{M_{{\rm
Li}^a}^{(\alpha)}}~+~\frac{ \Delta M_{{\rm Li}^b}^{(\alpha)}}{M_{{\rm
Li}^b}^{(\alpha)}} \right) \left( u_\infty^{\rm Li} - u_0^{\rm Li} \right),
\end{equation}
and that the analogous shift in the equilibrium distance $r_e$ is
\begin{equation}  \label{eq:dRe}
\delta r_e^{(\alpha)} ~=~ r_e^{(\alpha)} - r_e^{(1)} ~=~ - \left ( \frac{\Delta
M_{{\rm Li}^a}^{(\alpha)}}{M_{{\rm Li}^a}^{(\alpha)}}~+~\frac{ \Delta
M_{{\rm Li}^b}^{(\alpha)}}{M_{{\rm Li}^b}^{(\alpha)}} \right)
\frac{\widetilde S^{\,\prime}_{\rm ad}(r_e)}{\bar k},  
\end{equation}
in which $\,\bar{k}\,$ is the harmonic force constant at the potential
minimum in units $\,{\rm cm^{-1}\,\rm{\AA}^2}$, and 
\begin{equation}
\widetilde S^{\,\prime}_{\rm ad}(r_e) ~\equiv~ \left( \frac{d\widetilde S_{\rm ad}}{dr}
\right)_{r=r_e} =~~ \frac{(u_\infty - u_0)p_{\rm ad} + u_1\,q_{\rm ad}}
{ 2\,r_e}.  \label{eq:u1x}
\end{equation}
Similarly, the electronic isotope shift will be 
\begin{eqnarray}
\delta\left\{\Delta T_e^{(\alpha)}\right\} &=& \Delta T_e^{(\alpha)} - 
  \Delta T_e^{(1)} \label{eq:dTe} \\
&=& \left ( \frac{\Delta M_{{\rm Li}^a}^{(\alpha)}}{M_{{\rm
  Li}^a}^{(\alpha)}}~+~\frac{ \Delta M_{{\rm Li}^b}^{(\alpha)}}{M_{{\rm
  Li}^b}^{(\alpha)}} \right) \left[ u_0^{\rm Li}(1\,^3\Sigma_g^+) - 
  u_0^{\rm Li}(a\,^3\Sigma_u^+) \right].
  \nonumber
\end{eqnarray}
Note that in the present context, ${\mathfrak D}_e^{(1)}$ and $r_e^{(1)}$
are the values of the well depth and equilibrium distance of the MLR
potential for the reference isotopologue species, and are determined by
the DPF analysis.

Finally, as was pointed out by McAlexander {\em et al.},\cite{d:mcal96}
for the $1\,{^3\Sigma}_g^+$ state of Li$_2$, the dominant BOB contribution
to the rotationless potential at large $r$ has the form
\begin{equation}  \label{eq:dVna}
\Delta V_{\rm ad}^{(\alpha)} ~\simeq~ 2\,B^{(\alpha)}(r) ~=~ 
  2 \left(\frac{\hslash^2} {2\mu_\alpha\,r^2} \right),
\end{equation}
and since $\mu_\alpha$ for isotopic Li$_2$ is relatively small,
this behavior must be considered.  Following the approach of
Refs.\,\onlinecite{d:mcal96,d:vogt07,d:lero09}, we have chosen to treat
this term as a separate additive contribution to the effective interaction
potential for each isotopologue, which therefore takes on the form:
\begin{equation}   \label{eq:Vtot}
V_{\rm ad,tot}^{(\alpha)}(r)~ =~ V_{\rm MLR}(r)~+~ \Delta V_{\rm
ad}^{(\alpha)}(r) ~+~ 2\,B^{(\alpha)}(r)  ~~.
\end{equation}
As was pointed out by Vogt {\em et al.},\cite{d:vogt07} this $\Delta
V_{\rm ad}(r)$ term is readily incorporated into the Hamiltonian
by simply replacing the factor $[J(J+1)]$ in Eq.\,(\ref{eq:schr}) by
$[J(J+1)+2]$, and their approach was adopted here.  However, this means
that the overall $1\,^3\Sigma_g^+$--state well depth and equilibrium
distance are actually $\,{\mathfrak D}_e^{\rm tot}(c) = {\mathfrak
D}_e^{(\alpha)} - 2\,B^{(\alpha)}(r_e)\,$ and $\,r_e^{\rm tot}(c)=
r_e^{(\alpha)}+ 4\,B^{(\alpha)}(r_e)/(\bar k\,r_e)\,$, where ${\mathfrak
D}_e^{(\alpha)}$ and $r_e^{(\alpha)}$ are defined by Eqs.\,(\ref{eq:dDe})
and (\ref{eq:dRe}), and ${\mathfrak D}_e^{(1)}$ and $r_e^{(1)}$ are the
(fitted) reference-isotopologue MLR parameters for that state.

\vspace{-4mm}
\section{Potentials for the $a\,^3\Sigma_u^+$ and
$1\,^3\Sigma_g^+$ States of ${\rm Li}_2$ }
\vspace{-4mm}

\subsection{Data Set and Methodology }
\vspace{-4mm}

An overview of the experimental data used in this work is presented in
Table I.  Most of the data (2555 out of 2792 observations), came from
fluorescence experiments performed by Martin {\em et al.}\cite{d:mart88}
and Linton {\em et al.}\cite{d:lint89}~ That data set was enlarged by
inclusion of 137 $2\,^3\Pi_g-a\,^3\Sigma_u^+$ transitions of $^{7,7}$Li$_2$
taken from the study of Ref.\,\onlinecite{d:lint99}, which extended the
$a\,^3\Sigma_u^+$ vibrational range to $\,v=9$.  Finally, information
about levels lying very near dissociation is provided by the one
available PAS datum for the $\,v=10\,$ level of the $a\,^3\Sigma_u^+$
state,\cite{d:abra95a} and 99 PAS data for the higher levels of the
$1\,^3\Sigma_g^+$ state.\cite{d:abra95b,dataPAS}~ Throughout this study,
the upper levels of all transitions originating in the the $2\,^3\Pi_g$
state were represented by independent fitted term values.  A listing of
the experimental data used in the present analysis in included in the
Supplementary Data supplied to the journal's www archive

All of the DPF data-analysis fits described herein were performed using
the program \texttt{DPotFit}, which is freely available (with a manual)
for download from the www.\cite{d:DPotFit}~ The initial trial values of
the parameters $\beta_i$ required for those fits were generated by applying
the program \texttt{betaFIT} (also available from the www)\cite{d:betaFIT}
to sets of turning points obtained from preliminary versions of the analysis.

\vspace{-4mm}
\subsection{Model for Li$_2(a\,^3\Sigma_u^+)$ }
\vspace{-4mm}

The $a\,^3\Sigma_u^+$ state of Li$_2$ dissociates to yield two $S$-state
atoms, and ignoring hyperfine effects, there is no noteworthy interstate
coupling.  The theory of intermolecular forces therefore tells us that the
leading contributions to the long-range intermolecular potential should
consist of terms associated with (inverse) powers $\,m_i=\{6, 8, 10\}$.
The present analysis therefore represented the potential energy for this
species by an MLR potential incorporating the long-range tail function
\begin{equation}  \label{eq:uLRlower}
u_{\rm{LR}}^{\{a\,^3\Sigma_u^+\}} (r)~=~D_6(r)~\frac{C_6}{r^6}~+~
 D_8(r)~\frac{C_8}{r^8} ~+~ D_{10}(r)~\frac{C_{10}}{r^{10}} ~~,
\end{equation}
in which $D_m(r)$ are the modified Douketis-type damping functions of
Eq.\,(\ref{eq:DSdamp}), $\,\rho= 0.54$,\cite{d:lero11} and the dispersion
energy coefficients for this state were fixed at the values reported by Tang
\textit{et al.}\cite{d:tang09}~

In the initial work to determine an optimum model for this state, the PAS
data for the $1\,^3\Sigma_g^+$ state were ignored and all of the observed
levels of both the $1\,^3\Sigma_g^+$ and $2\,^3\Pi_g$ states were represented
by independent term values, so only a single potential energy function was
involved in the analysis.  Fits were then performed to a wide variety of
models corresponding to different choices for the order $N$ of the polynomial
in Eq.\,(\ref{eq:betapq}), and for the power $q$ and the reference distance
$r_{\rm ref}$ of Eq.\,(\ref{eq:ypq_ref}).  As was pointed out in \S\,II.B,
the power $p$ must be larger than the difference between the largest and
smallest powers of the terms contributing to Eq.\,(\ref{eq:uLRlower}),
so it was fixed as $\,p=5$.

Figure 5 summarizes results obtained for six families of potential function
models of this type.  For any given values of $q$ and $r_{\rm ref}$,
the fits will always converge when the polynomial order $N$ becomes
sufficiently large.  This point is illustrated by the convergence of
the four families of solid triangular points shown in the lower panel of
Fig.\,5, which correspond to models with $\,q=3\,$ and $N$ increasing from
two to five.  The results for $\,N=3\,$ with $\,q=2\,$ and 4 (open square
and round points, respectively) show that the optimum value of $r_{\rm
ref}$ will depend on other features of the model, but convergence to the
same limiting quality-of-fit $\overline{dd}$ and essentially the same
limiting values of physically interesting parameters such as ${\mathfrak
D}_e$ will be achieved in any case.  The upper panel of Fig.\,5 shows the
fitted values of ${\mathfrak D}_e$ for the various models considered in
the lower panel.  Those results show that the fitted value of ${\mathfrak
D}_e$ may vary considerably from one model to another, but at the $r_{\rm
ref}$ value where $\overline{dd}$ approaches its minimum value, {\em all}
models converge to essentially the same value of ${\mathfrak D}_e$.

All of the fits that are summarized by Fig.\,5 used all of the 2693
`optical' data for the two isotopologues $^{7,7}$Li$_2$ and $^{6,7}$Li$_2$,
plus the one PAS datum which had been reported for the
$a\,^3\Sigma_u^+$ state, while the 395 term values of the observed levels
of the $1\,^3\Sigma_g^+$ state and 20 term values for the $2\,^3\Pi_g$
state were treated as free parameters.  As might be expected when dealing
with light atoms such as Li, BOB effects are not negligible in this system.
In particular, it was found that allowing for one non-zero term $u_0^{\rm
Li}$ in the  expression for the `adiabatic' correction radial strength
function of Eq.\,(\ref{eq:Sad}) reduced the value of $\overline{dd}$
by 2.4\%; however, freeing a second coefficient ($u_1^{\rm Li}$) only
reduced $\overline{dd}$ by an additional 0.08\%, and the resulting value
of $u_1^{\rm Li}$ had an uncertainty of greater than 100\%.  Similarly,
allowing one centrifugal BOB parameter $t_1^{\rm Li}$ to vary led to
reductions in $\overline{dd}$ of less that 0.05\%.  As a result, the model
actually used to obtain the results summarized in Fig.\,5 included only
the one free BOB parameter $u_0^{\rm Li}$.

All else being equal, the ``best'' model for a given system is the one
which achieves an optimum quality of fit (lowest $\overline{dd}$) with
the smallest number of free parameters.  When more than that minimum
number of parameters are used, the additional degrees of freedom in
parameter space will not be strongly constrained by the data, and the
possibility of problems in the extrapolation regions tends to increase.
On this basis we choose the M3LR$_{5,3}^{8.0}(3)$ model corresponding
to the `gradiant-shaped' black points at $r_{\rm ref}= 8.0\,$\AA\ on
Fig.\,5 as our preferred model.  To four decimal places, increasing $N$
by one had no effect on the associated value of $\,\overline{dd}=0.7069$.
While the analysis described above led to our determination of optimal
models for the $a\,^3\Sigma_u^+$--state potential and BOB function,
the associated parameters were also allowed to vary freely in the global
two-state fit which simultaneously determined potential energy functions
for the $1\,^3\Sigma_g^+$ state.  Hence, they will be reported later.

One final component of this discussion is an illustration of the remarkably
robust extrapolation properties of the MLR potential form.  In the initial
stages of this study, only the optical $1\,^3\Sigma_g^+ -a\,^3\Sigma_u^+$
data which spanned vibrational levels $\,v(a)=0-7\,$ were considered
in the analysis.  The highest of the associated $a\,^3\Sigma_u^+$-state
levels is bound by 26 cm$^{-1}$ for $^{7,7}$Li$_2$ and by 16 cm$^{-1}$
for $^{6,6}$Li$_2$.  Nonetheless, the optimal MLR potential obtained
from that analysis was an M3LR$_{5,3}^{8.0}(r)$ function whose well
depth of $\,{\mathfrak D}_e= 333.79(1)~{\rm cm}^{-1}$ is very close to
the value $333.76(1)~{\rm cm}^{-1}$ yielded by the analysis of the full
$a\,^3\Sigma_u^+$--state data set (see Fig.\,5).  Moreover, the $\,v(a)=10\,$
binding energy predicted by that potential was 0.4222 cm$^{-1}$, which is
remarkably close to the measured PAS value\cite{d:abra95a} of $0.4160(\pm
0.0013)$ cm$^{-1}$.  Thus, a DPF analysis using an MLR potential with
a good multi-term theoretical $u_{\rm LR}(r)$ seems capable of yielding
quite reliable extrapolations to predict both the distance from the highest
observed level to dissociation and the number and energies of unobserved
higher levels.

\vspace{-4mm}
\subsection{Model for Li$_2(1\,^3\Sigma_g^+)$ and Results of the Two-State
Analysis }
\vspace{-4mm}

Following the discussion of \S\,II.C, the potential energy function
for the $1\,^3\Sigma_g^+$ state of Li$_2$ was represented by an MLR
potential whose long-range tail was defined by $u_{\rm LR}^{\rm ret}(r)$
of Eq.\,(\ref{eq:uLRfinal}).  As usual, fits were performed using models
corresponding to a variety of values of the exponent polynomial order $N$,
of the power $q$ defining the polynomial expansion variable, and of the
reference distance $r_{\rm ref}$ of Eq.\,(\ref{eq:ypq_ref}).  Since the
inverse-power terms contributing to $u_{\rm LR}^{\rm ret}(r)$ have
powers $\,m_i=\{3, 6, 8\}$, the power $p$ defining the radial variables
$y_p^{\rm eq}(r)$ and $y_p^{\rm ref}(r)$ of Eqs.\,(\ref{eq:VMLR}) and
(\ref{eq:ypq_ref}) was fixed at $\,p=6~(> 8-3)$, and while $C_6^\Sigma$
and $C_8^\Sigma$ were held fixed at the theoretical values of Tang
{\em et al.},\cite{d:tang09} $C_3^\Sigma$ was treated as a free
parameter.  All of these fits treated the full range of data for the
$1\,^3\Sigma_g^+$--$a\,^3\Sigma_u^+$ system, and while the potential for
the $a\,^3\Sigma_u^+$ state was represented by the M3LR$_{5,3}^{8.0}(r)$
model described in \S\,III.B, its parameters were also free variables in
these fits.

Figure 6 summarizes results for six families of $1\,^3\Sigma_g^+$--state
PECs: those for exponent polynomial orders $\,N=6-9$ with $\,q=3\,$
being represented by solid points, while those for polynomials orders
$\,N=8-9\,$ with $\,q=4\,$ (bottom panel only) are shown as open points.
As is expected, the fact that the 395 observed $1\,^3\Sigma_g^+$--state
level energies are now constrained to be eigenvalues of a potential function
rather than being free fitting parameters means that the $\overline{dd}$
values associated with the best of these fits are somewhat larger than
those for the $a\,^3\Sigma_u^+$--state analyses summarized in Fig.\,5.

As has been the case in other treatments of this type, for
any reasonable values of $q$ and $r_{\rm ref}$ the fits
converge to essentially the same optimum $\overline{dd}$
value when the polynomial order $N$ becomes sufficiently
large.\cite{d:lero09,d:coxo10,d:lero11,d:lero10a,d:hend11} A manual
optimization of $r_{\rm ref}$ is undertaken in order to determine a `best'
model, which is defined as one which: (i) gives a good fit to all data,
(ii) is defined by the smallest number of free parameters, and (iii)
has no unphysical behavior in the extrapolation regions.  As is usually
the case, models with larger $q$ values (here $\,q\!=\!4$) require a
higher-order polynomial to achieve a given quality of fit.  While not
shown, the $\overline{dd}$ values for $\,N=10$, $\,q=4\,$ models with
$\,r_{\rm ref}= 3.4-3.6\,$ are essentially identical to those for $\,N=9$,
$\,q=3\,$ models with the same $r_{\rm ref}$ (solid round points), but are
bigger at larger and smaller $r_{\rm ref}$ values.  Models with $\,q=1\,$
or 2 tended to have inflection points on the short-range repulsive wall,
even for cases with fairly large $r_{\rm ref}$ values.

The results in the two upper panels of Fig.\,6 show that the values of
${\mathfrak D}_e$ and $C_3^\Sigma$ yielded by fits to models with $\,N=9$,
$\,q=3\,$ vary relatively slowly with $r_{\rm ref}$, and that for $r_{\rm
ref}$ values which give small $\overline{dd}$ values, other types of models
yield very similar results.  For both of these properties the analogous
results for $\,q=4\,$ models were much more strongly model-dependent.
We therefore chose the $\,N=9$, $\,q=3\,$ potential with $\,r_{\rm ref}=
3.6\,$\AA\ as our recommended model for this state.

The error bars shown in Fig.\,6 are the 95\% confidence limit uncertainties
in the parameter yielded by the non-linear least-squares fits.  Although the
binding energies of the highest observed levels for the $1\,^3\Sigma_g^+$
state are even smaller than was the case for the $A\,^1\Sigma_u^+$
state,\cite{d:abra95b}  the uncertainties in the fitted $C_3^\Sigma$ values
obtained here ($\gtrsim 75\,{\rm cm^{-1}\,\AA^3}$) are an order of magnitude
larger than the analogous uncertainties yielded by the $A-X$ analysis of
Ref.\,\onlinecite{d:lero09}.  It is not clear why this should be the case,
other than the fact that the data gap from $\,v(1\,^3\Sigma_g^+)=8\,$ to 61
for $^{7,7}$Li$_2$ or to 55 for $^{6,6}$Li$_2$ may be expected to introduce
additional uncertainty into the analysis of the limiting near-dissociation
behavior.  However, the difference between the $C_3^\Sigma$ value implied
by the present analysis and that determined from the $A-X$ analysis of
Ref.\,\onlinecite{d:lero09} (dash-dot-dot line in the uppermost panel of
Fig.\,6) is significantly larger than the mutual uncertainties.  Moreover,
repeating the present analysis with $C_3^\Sigma$ fixed at the value yielded
by the $A-X$ analysis ($357\,829(\pm 8)~{\rm cm^{-1}\,\AA^3}$) increased the
overall value of $\overline{dd}$ by 0.8\%, and increased $\overline{dd}$ for
the PAS data by a massive 21\%!  It may be that a combined 5-state analysis
of the data sets for the two cases will resolve this discrepancy, but that
is beyond the scope of the present work.  Thus, our recommended model for
the $1\,^3\Sigma_g^+$ state is an M3LR$_{6,3}^{3.6}(9)$ potential with
$u_{\rm LR}(r)$ defined by Eq.\,(\ref{eq:uLRfinal}), and with $C_3^\Sigma$
determined from the fit.

The parameters defining our recommended models for the $a\,^3\Sigma_u^+$
and $1\,^3\Sigma_g^+$  states of Li$_2$ are listed in Table II.  The fact
that the uncertainties in the values of ${\mathfrak D}_e$ and $r_e$ are
an order of magnitude larger for the $1\,^3\Sigma_g^+$ state is to be
expected, both because of the 5200 cm$^{-1}$ data gap for $\,v>7$, and
because the fact that its lowest observed level is $\,v=1\,$ means that
the extrapolation to the potential minimum is much longer for this case.
As for the $a\,^3\Sigma_u^+$ state, obtaining a good combined-isotopologue
fit required the introduction of BOB corrections in the effective adiabatic
potential function for the $1\,^3\Sigma_g^+$ state.  As shown in Table II,
our recommended model includes a third-order polynomial expression for the
`adiabatic' correction radial strength function $\widetilde S_{\rm ad}^{\rm
Li}(r)$.  Increasing this polynomial order further or allowing for a non-zero
centrifugal BOB function yielded no significant improvement in the quality
of fit, while reducing this polynomial order by one or two terms increased
the $\overline{dd}$ value for the fit by 1.6\% or 3.8\%, respectively.

\vspace{-4mm}
\subsection{BOB Functions and Isotope Effects }
\vspace{-4mm}

The radial strength functions defining the effective adiabatic BOB
correction to the potential energy functions for both states are
shown in Fig.\,7.  Since $^{7,7}$Li$_2$ was chosen as the reference
isotopologue, Eq.\,(\ref{eq:dDe}) shows that the fact that $\,\widetilde
S_{\rm ad}^{\{a\,^3\Sigma_u^+\}}(r_e^{\{a\,^3\Sigma_u^+\}}) - \widetilde
S_{\rm ad}^{\{a\,^3\Sigma_u^+\}}(\infty) = u_0^{\{a\,^3\Sigma_u^+\}} -
u_\infty^{\{a\,^3\Sigma_u^+\}} = u_0^{\{a\,^3\Sigma_u^+\}}\,$ is positive
means that ${\mathfrak D}_e^{\{a\,^3\Sigma_u^+\}}$ is (slightly) larger for
$^{6,6}$Li$_2$ than for $^{7,7}$Li$_2$.  The upper curve in Fig.\,7 shows
that the $\widetilde S_{\rm ad}^{\{1\,^3\Sigma_g^+\}}(r)$ contribution to
the isotopologue dependence of ${\mathfrak D}_e^{\{1\,^3\Sigma_g^+\}}$ is
also positive (and much larger).  However, the isotopologue dependence of
the $2\,B^{(\alpha)}(r)$ contribution to the effective adiabatic potential
(see Eq.\,(\ref{eq:Vtot})) makes a negative ($-0.171~{\rm cm}^{-1}$)
contribution to the difference $\,{\mathfrak D}_e^{\rm tot(6,6)} -
{\mathfrak D}_e^{{\rm tot}(7,7)}$ for the $1\,^3\Sigma_g^+$ state, and
turns out to be the dominant term.

The results presented in Table II were obtained from an analysis which
treated $^{7,7}$Li$_2$ as the reference isotopologue, and the first
row of Table III presents characteristic properties of the resulting
$a\,^3\Sigma_u^+$ and $1\,^3\Sigma_g^+$ potential energy functions
for that species.  Note that the values of $\mathfrak{D}_e^{\rm
tot}(1\,^3\Sigma_g^+)$ and $r_e^{\rm tot}(1\,^3\Sigma_g^+)$ were obtained
after combining the MLR potential with the additive adiabatic correction
term of Eq.\,(\ref{eq:dVna}).  The next two rows of this table then show,
respectively, the isotopic changes in and the resulting values of these
quantities for $^{6,6}$Li$_2$, as implied by the BOB correction functions
${\widetilde S}_{\rm ad}(r)$ and (for the $1\,^3\Sigma_g^+$ state) the
$2\,B^{(\alpha)}(r)$ term (see Eqs.\,(\ref{eq:dDe})--(\ref{eq:Vtot})).  Of
course it is equally feasible to perform the overall analysis using
$^{6,6}$Li$_2$ as the reference isotopologue, and the last row of Table
III shows the properties of that isotopologue obtained in that more
direct manner.  It is reassuring to see that within the uncertainties,
the results in the last two rows of this table agree with one another.

Of course it is simpler to work with potential functions that do not
require the addition of separate adiabatic correction functions $\Delta
V_{\rm ad}^{(\alpha)}(r)$.  Hence, for the convenience of those interested
primarily in the minor isotopologue $^{6,6}$Li$_2$, a version of Table II
for the case in which this species was used as the reference isotopologue is
included in the Supplementary Data supplied to the journal's www archive.

\section{Discussion and Conclusions}

A combined-isotopologue DPF analysis of 2692 optical data for the
$1\,^3\Sigma_g^+-a\,^3\Sigma_u^+$ and $2\,^3\Pi_g-a\,^3\Sigma_u^+$ band
systems of $^{7,7}$Li$_2$ and $^{6,6}$Li$_2$, together with 99 PAS data
for the $1\,^3\Sigma_g^+$ state and one for the ${a\,^3\Sigma_u^+}$ state,
has yielded analytic potential energy functions for the $1\,^3\Sigma_g^+$
and $a\,^3\Sigma_u^+$ electronic states which (on average) explain all of
those data within the experimental uncertainties ($\,\overline{dd}= 0.789$).
The present potential energy function for the $a\,^3\Sigma_u^+$ state
of Li$_2$ is one of the most accurate ground-triplet-state potentials
determined for any alkali-atom pair.  The resulting scattering lengths for
$^{7,7}$Li$_2$ and $^{6,6}$Li$_2$ are $\,a_{\rm SL}= -14.759(9)$\,\AA\ and
$-1906(50)$\,\AA, respectively, where the uncertainties were estimated
by repeating the overall analysis with $C_6^{\{a\,^3\Sigma_g^+\}}$
increased/decreased by 0.01\% from the recommended values of Tang {\em et
al.}\cite{d:tang09}~ This 0.01\% is a factor of 3 larger than the $C_6$
uncertainty reported in Ref.\,\onlinecite{d:tang09}.  The uncertainty in
$a_{\rm SL}$ is much larger for $^{6,6}$Li$_2$ than for $^{7,7}$Li$_2$
simply because all else being equal, scattering lengths that are very
large in magnitude are much more sensitive to details of the potential
energy function.  Listings of band constants ($G_v$, $B_v$, $D_v$,
$H_v$, etc.) calculated from this potential for all bound levels of the
$a\,^3\Sigma_u^+$ state for all three Li$_2$ isotopologues are included
with the Supplementary Data supplied to the journal's www archive.

For the $1\,^3\Sigma_g^+$ state, the present analysis has provided
an analytic potential energy function which very robustly
bridges the $\sim 5200~{\rm cm}^{-1}$ gap shown in Fig.\,1,
between the fluorescence measurement domain $\,v= 1-7\,$ and the
PAS data for $\,v\geq 56\,$ for $^{6,6}$Li$_2$ and $\,\geq 62\,$
for $^{7,7}$Li$_2$.  To illustrate this interpolation behavior,
Fig.\,8 plots calculated properties of our recommended potential for
$^{7,7}$Li$_2$ in the manner suggested by near-dissociation theory
(NDT).\cite{d:lero70c,d:lero70d,d:lero72a,d:lero80a,d:lero10a}~ In
particular, NDT predicts that for vibrational levels lying near the
dissociation limit of a potential whose limiting long-range behavior is
defined by an attractive $C_3/r^3$ interaction energy, the $\nicefrac{1}{6}$
power of the binding energy $({\mathfrak D} - E_v)$, the $\nicefrac{1}{5}$
power of the vibrational level spacing $\Delta G_{v+\nicefrac{1}{2}}$,
and the $\nicefrac{1}{4}$ power of the inertial rotational constant $B_v$,
should all be linear functions of $v$, with slopes determined by the value
of the $C_3$ coefficient.  The solid triangular points in Fig.\,8 represent
the experimental data, while the open round points are our predictions
for the `no-data' regions.  The dash-dot-dot lines in Fig.\,6 are the
limiting NDT slopes implied by the fitted $C_3^\Sigma$ value of Table II.
The deviation from this behavior at very high $v$ reflects the fact that the
$3\!\times\!3$ interstate coupling reduces the magnitude of the effective
$C_3$ coefficient in the limiting region by a factor of $\nicefrac{1}{3}$
(see Fig.\,4) as one approaches the limit.  Calculated band constants for all
bound levels of all three Li$_2$ isotopologues in this state have been placed
in the journal's Supplementary Data archive.  

It is noteworthy that predictions generated from a variety of other MLR
potential models (i.e., models defined by different $N$ or $q$ values)
which yield good fits to the data are identical on the scale of Fig.\,8.
This model-independent bridging of a data-gap spanning 73\% of the well
depth is a remarkable illustration of the robustness of the MLR potential
function form.  The ability of this function to readily incorporate the
effect of two-state\cite{d:lero09} or three-state (present work) coupling
in the long-range region is a further demonstration of its capabilities.
A {\sc Fortran} subroutine for generating the recommended potentials is
one of the items placed in the journal's Supplementary Data archive.

One puzzle left by this work is the discrepancy between the
value of $C_3^\Sigma$ for interacting Li($^2P$) + Li($^2S)$ atoms
determined in the present analysis ($3.575\,57(78)\times 10^5~{\rm
cm^{-1}\,\AA^3}$) and those obtained in the $A(^1\Sigma_u^+)$--state
analysis of Ref.\,\onlinecite{d:lero09} ($3.578\,29(7)\times 10^5~{\rm
cm^{-1}\,\AA^3}$) or from the recent theoretical calculations of Tang {\em et
al.},\cite{d:tang09}  ($3.578\,108\,9(7)\times 10^5~{\rm cm^{-1}\,\AA^3}$).
While small on an absolute scale, this 0.076\% discrepancy is much larger
than the estimated uncertainties, and repeating our overall analysis with
$C_3^\Sigma$ fixed at the $A$-state value from Ref.\,\onlinecite{d:lero09}
yielded a distinctly poorer quality fit, especially for the PAS data.
It may be that a combined five-state analysis of all of the data considered
here with those used in the $A-X$ analysis of Ref.\,\onlinecite{d:lero09}
will shed light on this question, but that will have to await future work.

\subsection*{Acknowledgements}

We are very grateful to Dr.\ Amanda Ross for stimulating discussions
which brought this problem to our attention, and to Professor F.R.W.\
McCourt for helpful discussions.  This research has been supported by
the Natural Sciences and Engineering Research Council of Canada.


\newpage

\begin{table}[th]
\begin{description}
\item[Table I]
Summary of experimental data used in the present work.
\end{description}
\vspace{-6mm}
\begin{center}
\begin{tabular}{l l c c c c c c}
\hline\hline
isotop. & type & unc.\ (cm$^{-1}$) & $~v(2\,^3\Pi_g)~$ & $~v(1\,^3\Sigma_g^+)~$
   & $~v(a\,^3\Sigma_u^+)~$ & $~^\#$\,data~ & source  \\
\hline
$^{7,7}$Li$_2$  &  $1\,^3\Sigma_g^+$ emission & $0.01$ & --- &  $1-7$  & $0-7$ 
   & 1279 & Ref.\,\onlinecite{d:mart88} \\
 &  $2\,^3\Pi_g$ emission & $0.005-0.01$  & $1-2$ & --- & $0-9$ & 137
    & Ref.\,\onlinecite{d:lint99}  \\
 &  PAS($1\,^3\Sigma_g^+$)    & $0.0043-0.00073$ & --- & $63-90$ & --- & 30 
    & Ref.\,\onlinecite{d:abra95b} \\ 
 &  PAS($a\,^3\Sigma_u^+$)    & $0.0013$ & --- & --- & $10$ &  1
   & Ref.\,\onlinecite{d:abra95a}  \\
\noalign{\medskip}
$^{6,6}$Li$_2$  &  $1\,^3\Sigma_g^+$ emission & 0.01 & --- & $1-7$ & $0-7$ 
   &  1276 & Refs.\,\onlinecite{d:lint89}  \\
 &  PAS($1\,^3\Sigma_g^+$)    & ~~$0.00110-0.01067 $~~  & --- & $56-84$  &  --- 
   &  69 & Ref.\,\onlinecite{d:abra95b}  \\
\multicolumn{2}{c}{\textbf{Overall}} &  ~~  & $1-2$ & $1-90$  & $0-10$ & 2792 &  \\
\noalign{\medskip}
\hline\hline
\end{tabular}
\end{center}
\label{tab:data}
\end{table}

\begin{table}[th]
\renewcommand \baselinestretch{1.1}
\normalsize
\begin{description}
\item[Table II]
Parameters defining the recommended MLR potentials and BOB functions for
the $a\,^3\Sigma_u^+$ and $1\,^3\Sigma^+_g$ states of Li$_2$ obtained
using $^{7,7}$Li$_2$ as the reference isotopologue.  Parameters in square
brackets were held fixed in the fit, while numbers in round brackets
are 95\% confidence limit uncertainties in units of the last digits show.
The analysis used the $^7$Li $^2P_{1/2}\leftarrow 2S_{1/2}$ excitation energy
of 14903.648130 cm$^{-1}$ and $^2P_{3/2}\leftarrow 2P_{1/2}$ spin-orbit
splitting energy of 0.335338 cm$^{-1}$ from Ref.\,\onlinecite{d:sans95}~
Units of length and energy are \AA\ and cm$^{-1}$; the exponent expansion
coefficients $\beta_i$ are dimensionless, while the parameters $u_i$
defining the `adiabatic' BOB strength function of Eq.\,(\ref{eq:Sad})
have units cm$^{-1}$.

\end{description}
\vspace{-15mm}
\begin{center}
\[ \begin{array}{ c r@{.}l c r r@{.}l }
\hline\hline
\noalign{\smallskip}
&\multicolumn{2}{c}{a(^3\Sigma_u^+)} && & \multicolumn{2}{c}{c(1\,^3\Sigma^+_g)} \\
\noalign{\smallskip}
\hline
\noalign{\smallskip}
\mathfrak{D}_e  & \multicolumn{2}{c}{333.758\,(7)}  
          &~~ & & \multicolumn{2}{c}{7093.44\,(3)~~~~}  \\
r_e         & 4&17005\,(3)          &~~~ & ~~~~~~~    &  3&06514\,(9)  \\
C_6      &[\,6&7185\times 10^6\,]  & & C_3    &  3&57557\,(78)\times 10^5 \\
C_8      &[\,1&12629\times 10^8\,] & & C_6^\Sigma & [\,1&00054\times 10^7\,] \\
C_{10}   &[\,2&78683\times 10^9\,] & & C_8^\Sigma & [\,3&69953\times 10^8\,] \\
\rho_{\rm Li} & \multicolumn{2}{l}{$~~~[\,0.54\,] $} & & & 
\multicolumn{2}{l}{$~~~~N/A$}  \\
\{p,\,q\}   & \multicolumn{2}{l}{~~~\{5,\,3\} } 
        & & & \multicolumn{2}{l}{~~~\{6,\,3\} }  \\
r_{\rm ref} & \multicolumn{2}{l}{~~~~[\,8.0\,]} &&& \multicolumn{2}{l}{~~~~[\,3.6\,]} \\
\beta_0     & -0&51608              & & &     -1&6373863 \\
\beta_1     & -0&09782              & & &      0&29197 \\
\beta_2     &  0&1137               & & &     -0&55544 \\
\beta_3     & -0&0248               & & &     -0&2794 \\
\beta_4     & \multicolumn{2}{c}{$---$} & & & -1&5993 \\
\beta_5     & \multicolumn{2}{c}{$---$} & & & -0&673 \\
\beta_6     & \multicolumn{2}{c}{$---$} & & & -1&23 \\
\beta_7     & \multicolumn{2}{c}{$---$} & & & -1&29 \\
\beta_8     & \multicolumn{2}{c}{$---$} & & &  0&5 \\
\beta_9     & \multicolumn{2}{c}{$---$} & & &  2&6 \\
\noalign{\medskip}
\{p_{\rm ad},\,q_{\rm ad}\}  & \multicolumn{2}{l}{~~~\{6,\,6\}} 
                         & & & \multicolumn{2}{l}{~~~\{3,\,3\}} \\
u_0              &  0&059\,(11)                & & &   1&367\,(7) \\
u_1              & \multicolumn{2}{l}{~~~$---$ } & & & 2&7 \,(4) \\
u_2              & \multicolumn{2}{l}{~~~$---$ } & & &-1&3 \,(5)   \\
u_3              & \multicolumn{2}{l}{~~~$---$ } & & &-1&8 \,(4)   \\
u_\infty         & [0&0]                         & & & [1&055740]  \\
\noalign{\medskip}
\hline\hline
\end{array} \]
\end{center}
\end{table}

\begin{table}[th]
\renewcommand \baselinestretch{1.1}
\normalsize
\begin{description}
\item[Table II.B] (for Supplementary Data submission)

Parameters defining the recommended MLR potentials and BOB functions for
the $a\,^3\Sigma_u^+$ and $1\,^3\Sigma^+_g$ states of Li$_2$ obtained
using $^{6,6}$Li$_2$ as the reference isotopologue.  Parameters in square
brackets were held fixed in the fits, while numbers in round brackets are
the 95\% confidence limit uncertainties in units of the last digits show.
The analysis used the $^6$Li $^2P_{1/2}\leftarrow 2S_{1/2}$ excitation energy
of 14903.296792 cm$^{-1}$ and $^2P_{3/2}\leftarrow 2P_{1/2}$ spin-orbit
splitting energy of 0.335324 cm$^{-1}$ from Ref.\,\onlinecite{d:sans95}~
Units of length and energy are \AA\ and cm$^{-1}$; the exponent expansion
coefficients $\beta_i$ are dimensionless, while the parameters $u_i$
defining the `adiabatic' BOB strength function of Eq.\,(\ref{eq:Sad})
have units cm$^{-1}$.

\end{description}
\vspace{-15mm}
\begin{center}
\[ \begin{array}{ c r@{.}l c r r@{.}l }
\hline\hline
\noalign{\smallskip}
&\multicolumn{2}{c}{a(^3\Sigma_u^+)} && & \multicolumn{2}{c}{c(1\,^3\Sigma^+_g)} \\
\noalign{\smallskip}
\hline
\noalign{\smallskip}
\mathfrak{D}_e  & \multicolumn{2}{c}{333.778\,(7)}  
          &~~ & & \multicolumn{2}{c}{7093.54\,(3)~~~~}  \\
r_e         & 4&17001\,(3)      &~~~ & ~~~~~~~ & 3&06520\,(9)  \\
C_6      &[\,6&7190\times 10^6\,]  & & C_3     & 3&57549\,(78)\times 10^5 \\
C_8      &[\,1&12634\times 10^8\,] & & C_6^\Sigma & [\,1&00059\times 10^7\,] \\
C_{10}   &[\,2&78694\times 10^9\,] & & C_8^\Sigma & [\,3&69965\times 10^8\,] \\
\rho_{\rm Li} & \multicolumn{2}{l}{$~~~[\,0.54\,] $} & & & 
\multicolumn{2}{l}{$~~~~N/A$}  \\
\{p,\,q\}   & \multicolumn{2}{l}{~~~\{5,\,3\} } 
        & & & \multicolumn{2}{l}{~~~\{6,\,3\} }  \\
r_{\rm ref} & \multicolumn{2}{l}{~~~~[\,8.0\,]} &&& \multicolumn{2}{l}{~~~~[\,3.6\,]} \\
\beta_0     & -0&516086             & & &     -1&6373493 \\
\beta_1     & -0&09783              & & &      0&29191 \\
\beta_2     &  0&1136               & & &     -0&55591 \\
\beta_3     & -0&0249               & & &     -0&2788 \\
\beta_4     & \multicolumn{2}{c}{$---$} & & & -1&5931 \\
\beta_5     & \multicolumn{2}{c}{$---$} & & & -0&685 \\
\beta_6     & \multicolumn{2}{c}{$---$} & & & -1&306 \\
\beta_7     & \multicolumn{2}{c}{$---$} & & & -1&34 \\
\beta_8     & \multicolumn{2}{c}{$---$} & & &  0&6 \\
\beta_9     & \multicolumn{2}{c}{$---$} & & &  2&7 \\
\noalign{\medskip}
\{p_{\rm ad},\,q_{\rm ad}\}  & \multicolumn{2}{l}{~~~\{6,\,6\}} 
                         & & & \multicolumn{2}{l}{~~~\{3,\,3\}} \\
u_0              &  0&067\,(13)                & & &   1&595\,(8) \\
u_1              & \multicolumn{2}{l}{~~~$---$ } & & & 3&1 \,(5) \\
u_2              & \multicolumn{2}{l}{~~~$---$ } & & &-1&5 \,(5)   \\
u_3              & \multicolumn{2}{l}{~~~$---$ } & & &-2&1 \,(5)   \\
u_\infty         & [0&0]                         & & & [1&23141]  \\
\noalign{\medskip}
\hline\hline
\end{array} \]
\end{center}
\end{table}

\begin{table}[th]
\begin{description}
\item[Table III]
Properties of the recommended potential energy functions for the
$a\,^3\Sigma_u^+$ and $1\,^3\Sigma_g^+$ states of Li$_2$ (energies
in cm$^{-1}$ and lengths in \AA), with `{\em changes}' calculated using
Eqs.\,(\ref{eq:dDe})--(\ref{eq:Vtot}).  The first three rows correspond
to use of $^{7,7}$Li$_2$ as the reference isotopologue, while for the last
row it was $^{6,6}$Li$_2$.

\end{description}
\renewcommand \baselinestretch{1.5}
\normalsize
\vspace{-16mm}
\begin{center}
\[ \begin{array}{ c c r@{.}l r@{.}l r@{.}l r@{.}l r@{.}l }
\hline\hline
  &   & \multicolumn{4}{c}{a\,^3\Sigma_u^+~$state$} & \multicolumn{2}{c}{~~} & 
        \multicolumn{4}{c}{1\,^3\Sigma_g^+~$state$} \\
\cline{3-6} \cline{9-12}
$fit$ & $isot.$ & \multicolumn{2}{c}{\mathfrak{D}_e} 
  & \multicolumn{2}{c}{r_e} 
  & \multicolumn{2}{c}{\Delta T_e} 
  & \multicolumn{2}{c}{\mathfrak{D}_e^{\rm tot}} 
  & \multicolumn{2}{c}{r_e^{\rm tot}} \\
\hline
$2-isot$ & ^{7,7}{\rm Li}_2   &~~ 333&758(7) &  4&17005(3)   & ~~8144&989(43)~~ & 7092&417(33) & 3&06524(9)  \\
\multicolumn{2}{c}{\it change}&~~ 0&019(4)   & -0&000015(3)  & ~~  -0&265(5) ~~ & -0&067(2)    & 0&000075(1) \\
 & ^{6,6}{\rm Li}_2           &~~ 333&777(8) &  4&17003_5(3) & ~~8144&724(43)~~ & 7092&349(33) & 3&06532(9)  \\
\hline
$2-isot$ & ^{6,6}{\rm Li}_2   &~~ 333&778(7) &  4&17001(3)   & ~~8144&726(33)~~ & 7092&347(33) & 3&06532(9)  \\
\hline\hline
\end{array} \]
\end{center}

\end{table}

\clearpage

\begin{figure}[h!]
\begin{center}
\scalebox{0.80}{
\includegraphics*{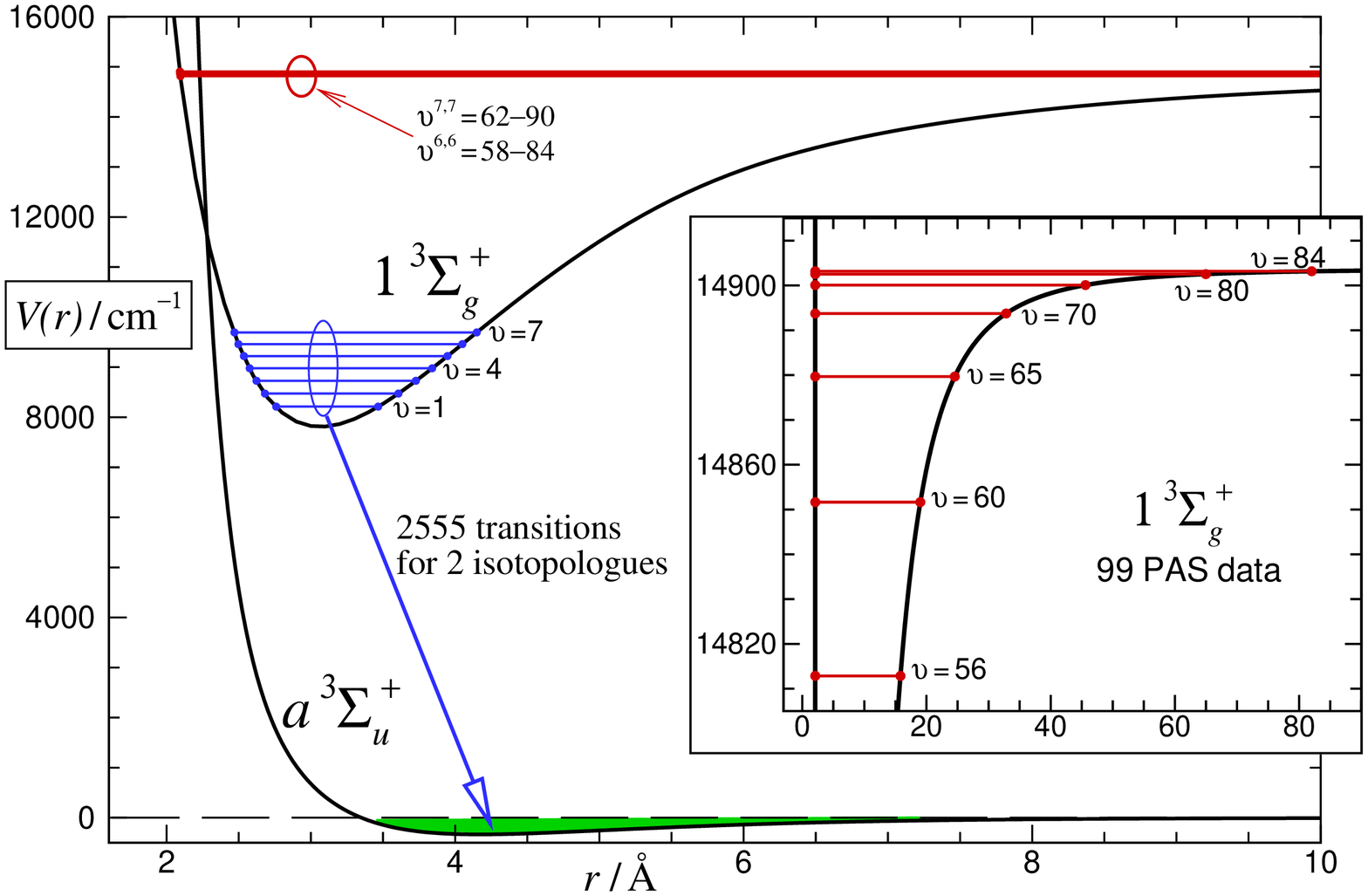} }
\end{center}
\begin{description}
\vspace{-8mm}
\item[Figure 1.]
Overview of the potential energy functions and observed vibrational levels of
the $1\,^3\Sigma_g^+$ and $a\,^3\Sigma_u^+$ states associated with the present
analysis.  The insert shows a fragment of the $1\,^3\Sigma_g^+$ state potential
at the energy range associated with the $^{6,6}$Li$_2$ PAS data.
\end{description}

\end{figure}

\begin{figure}[h!]
\begin{center}
\scalebox{0.80}{
\includegraphics{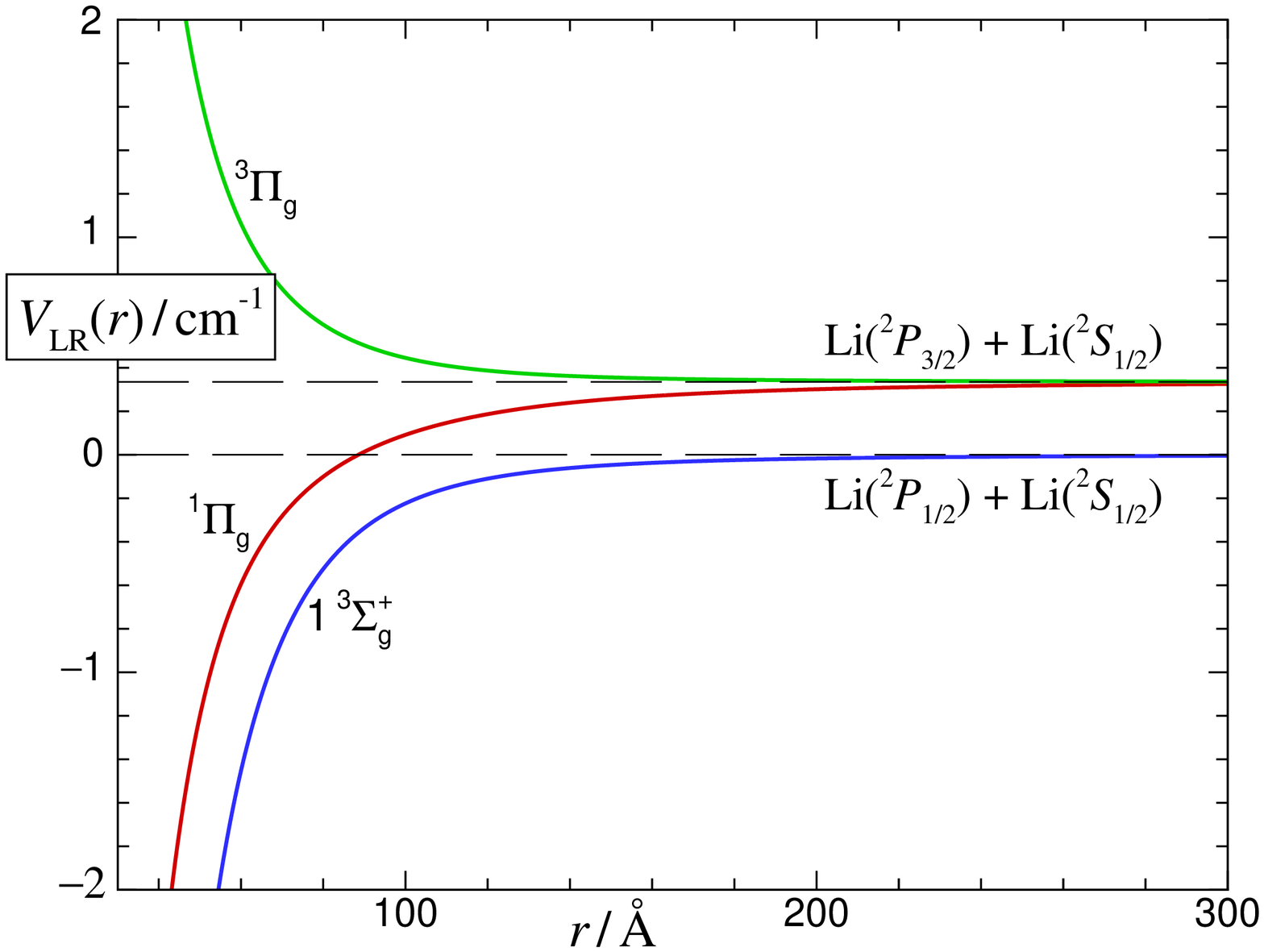} }
\begin{description}
\vspace{-4mm}
\item[Figure 2.]
Eigenvalues of Eq.\,(\ref{eq:matrix}) for the $1_g$ states of Li$_2$
dissociating to the Li$(2s\,^2S)+{\rm Li}(2p~^2P)$ limits, as generated
using the long-range coefficients of Tang {\em et al.}\cite{d:tang09}
\end{description}
\end{center}
\end{figure}

\begin{figure}[h!]
\begin{center}
\scalebox{0.8}{
\includegraphics{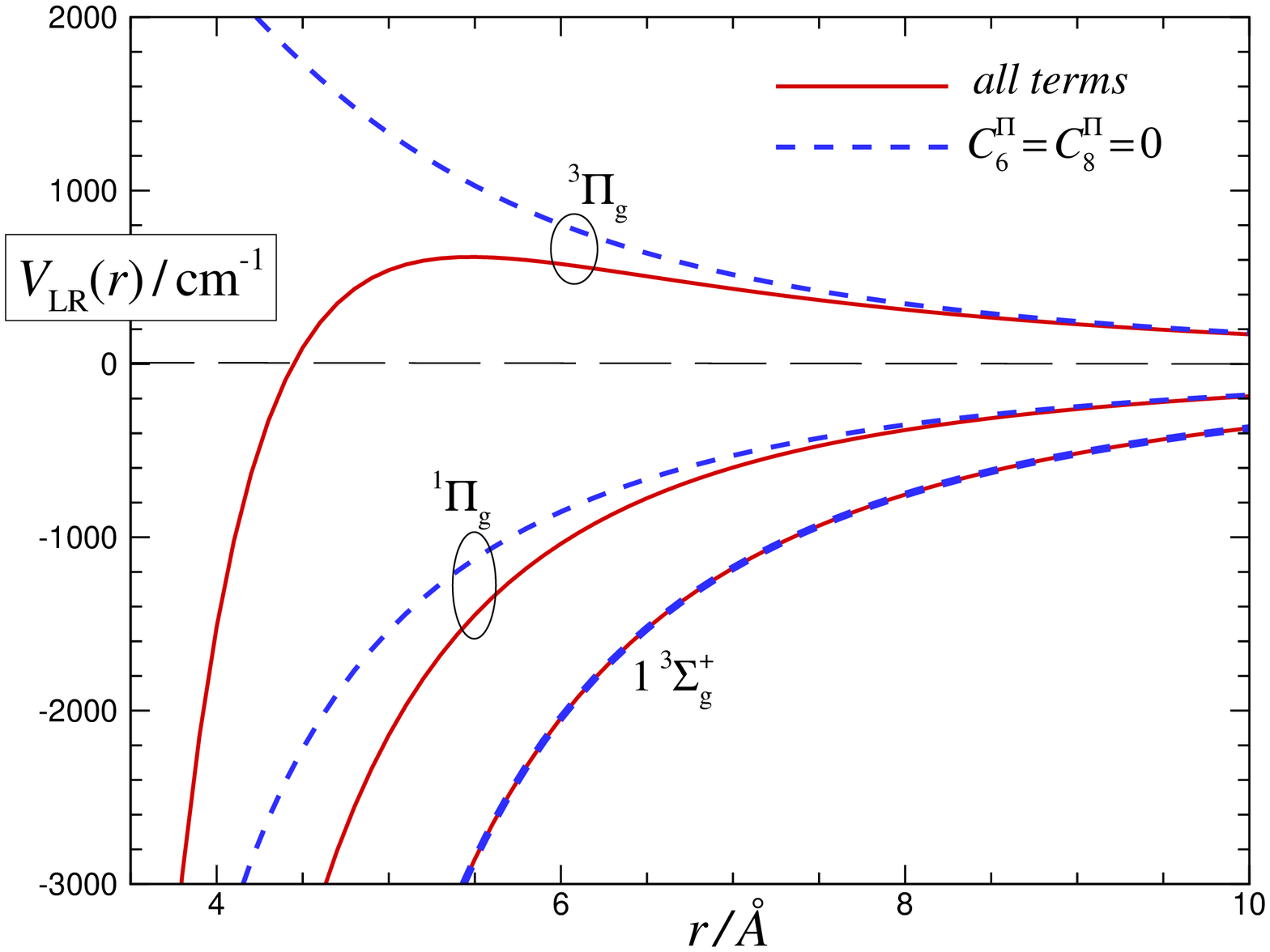}  }
\begin{description}
\vspace{-4mm}
\item[Figure 3.]
Illustration of the effect of setting  $\,C_6^\Pi=  C_8^{^1\Pi_g}=
C_8^{^3\Pi_g}= 0\,$ in Eq.\,(\ref{eq:matrix}).
\end{description}
\end{center}
\end{figure}

\begin{figure}[h!]
\begin{center}
\scalebox{0.95}{\hspace{-10mm}
\includegraphics{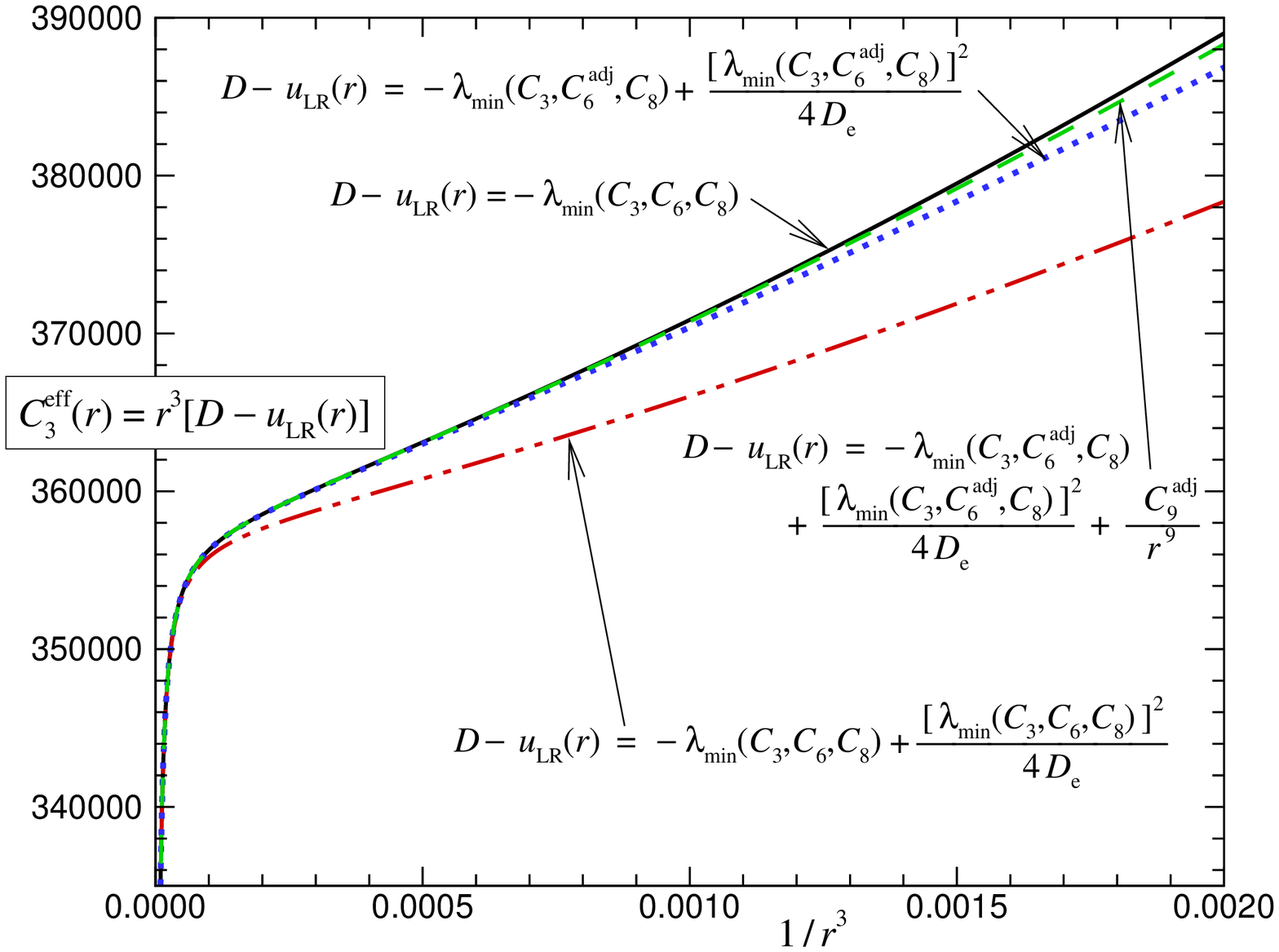}  }
\begin{description}
\vspace{-4mm}
\item[Figure 4]
Comparison of four representations of the long-range potential for the
$1\,^3\Sigma_g^+$ state of Li$_2$, with energies in cm$^{-1}$ and lengths in
\AA.
\end{description}
\end{center}
\end{figure}

\begin{figure}[h!]
\begin{center}
\vspace{-10mm}
\scalebox{0.95}{\hspace{-10mm}
\includegraphics{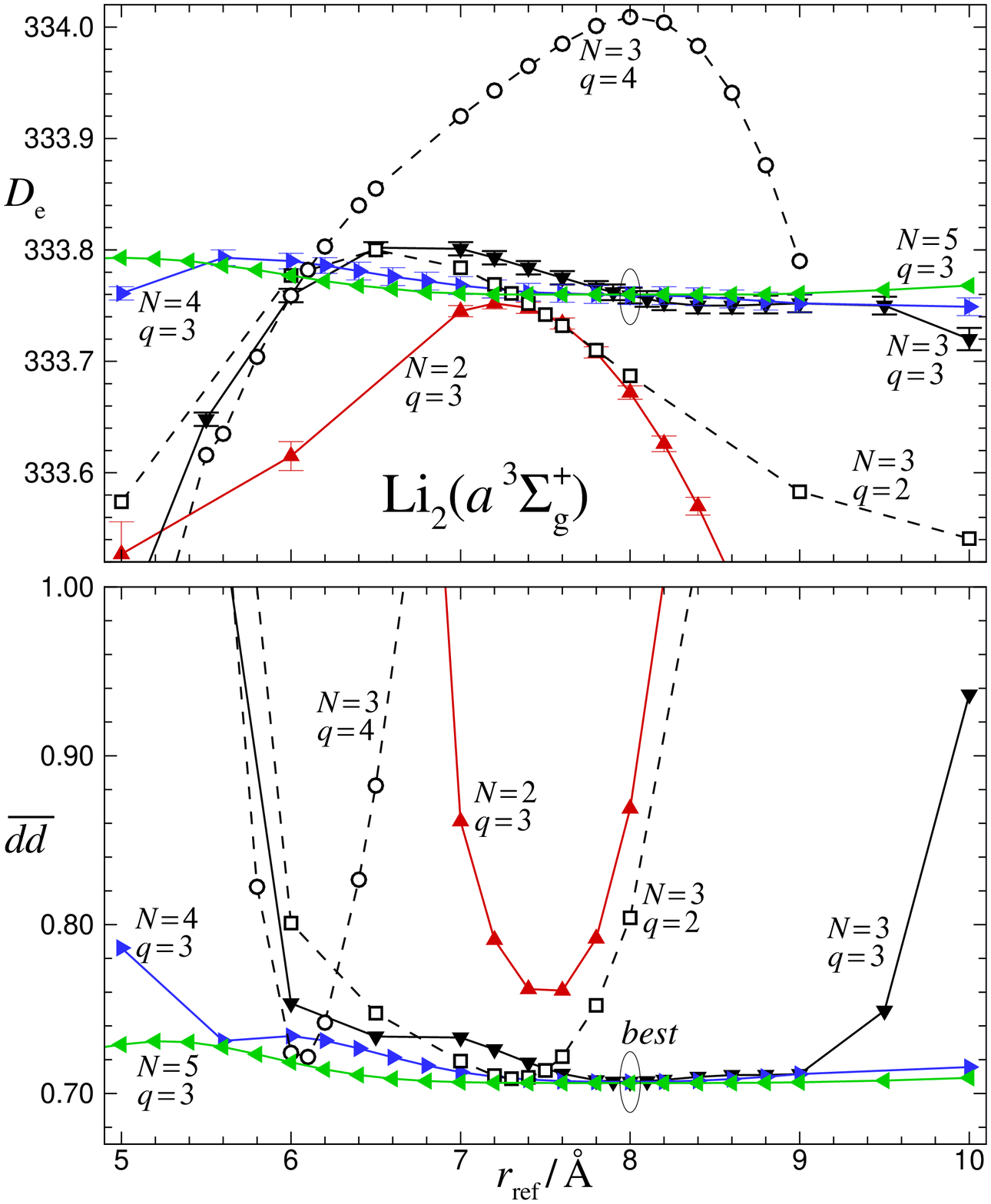}  }
\begin{description}
\vspace{-4mm}
\item[Figure 5]
Determination of the optimum model for the $a\,^3\Sigma_u^+$ state of Li$_2$.
\end{description}
\end{center}
\end{figure}

\begin{figure}[h!]
\begin{center}
\vspace{-10mm}
\scalebox{0.95}{\hspace{-10mm}
\includegraphics{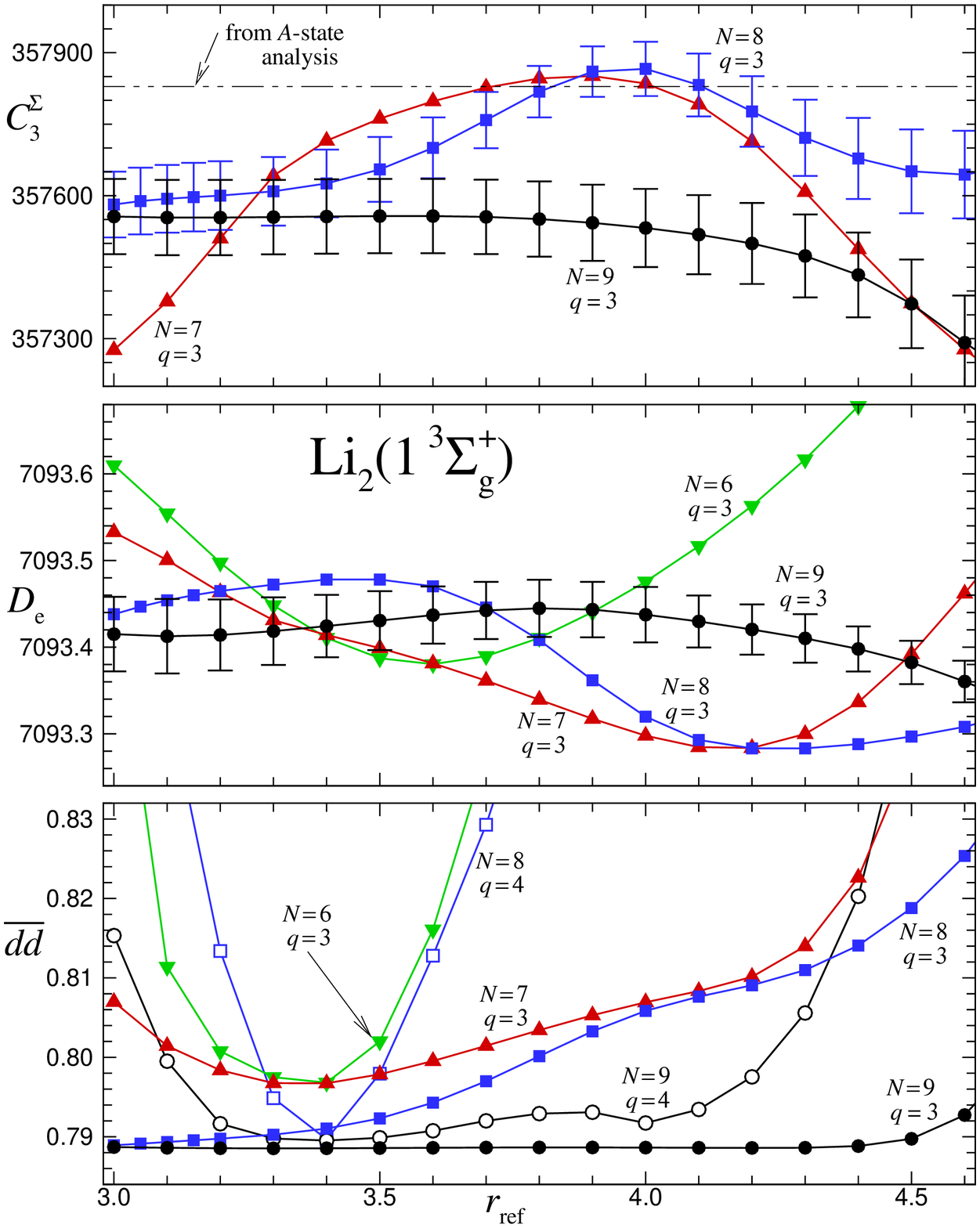}  }
\begin{description}
\vspace{-4mm}
\item[Figure 6]
Determination of the optimum model for the $1\,^3\Sigma_g^+$ state of Li$_2$.
\end{description}
\end{center}
\end{figure}

\begin{figure}[h!]
\begin{center}
\vspace{-10mm}
\scalebox{0.95}{\hspace{-10mm}
\includegraphics{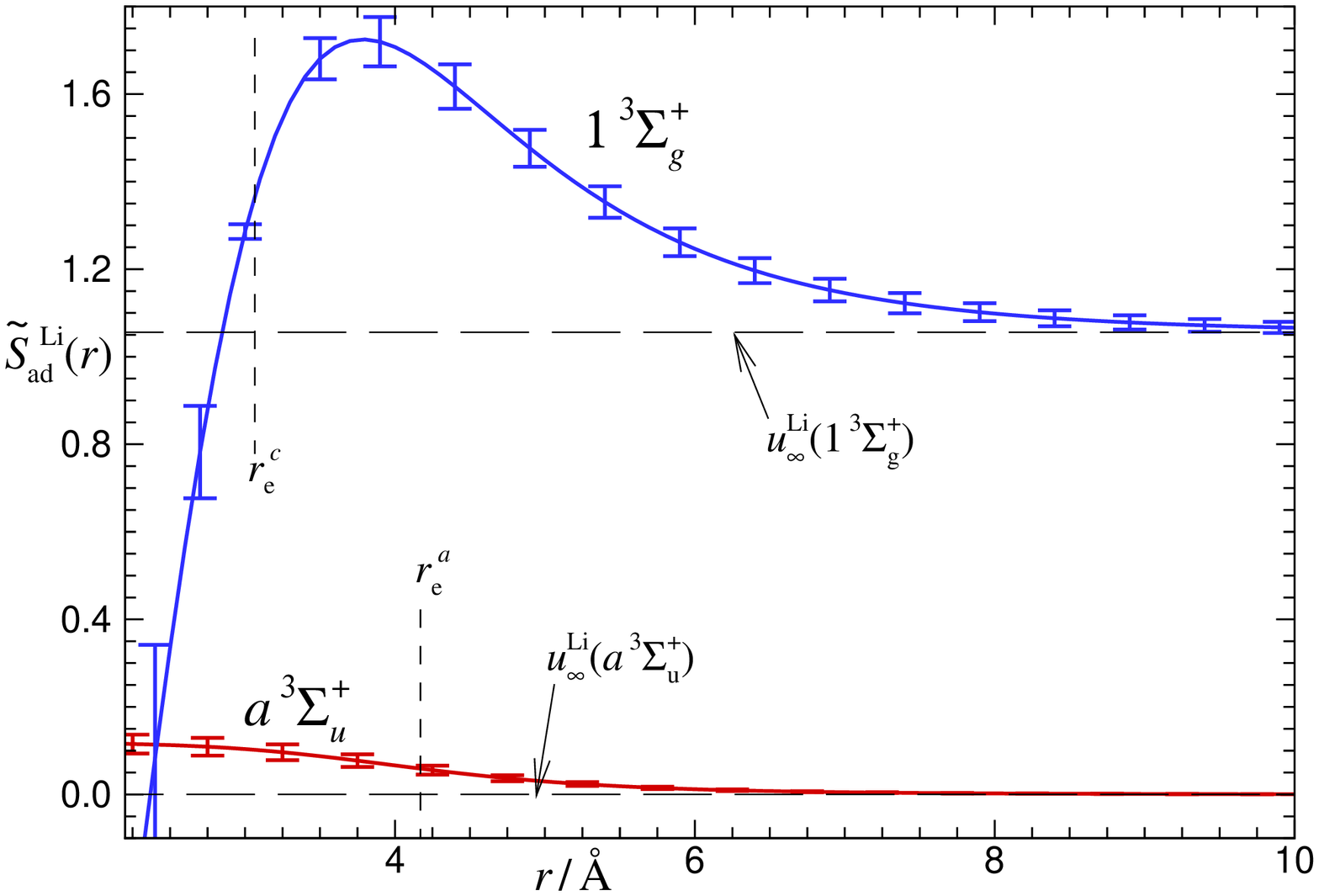}  }
\begin{description}
\vspace{-4mm}
\item[Figure 7]
BOB radial strength functions (in cm$^{-1}$) determined with $^{7,7}$Li$_2$ 
as the reference isotopologue. 
\end{description}
\end{center}
\end{figure}

\begin{figure}[h!]
\begin{center}
\vspace{-10mm}
\scalebox{0.95}{\hspace{-10mm}
\includegraphics{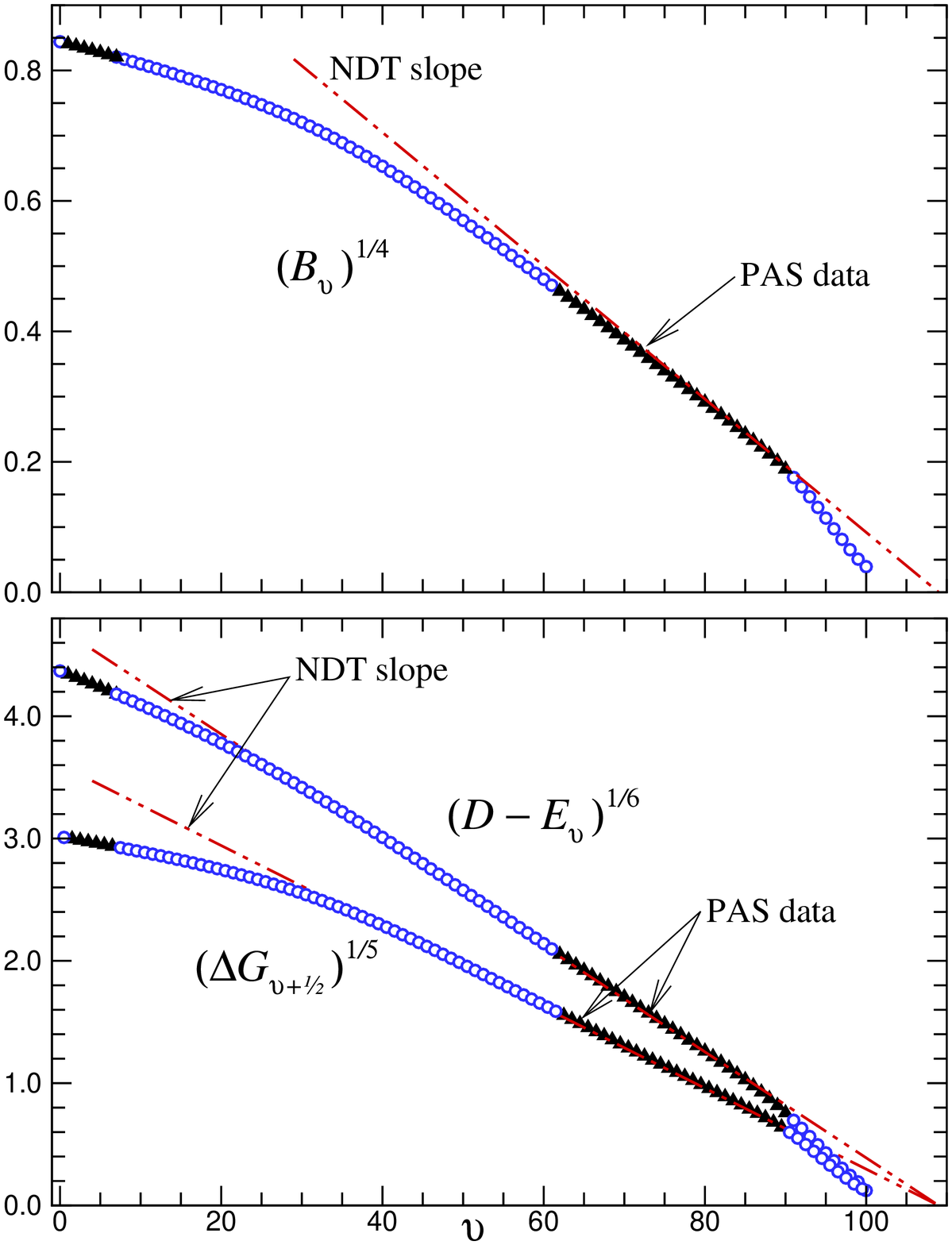}  }
\begin{description}
\vspace{-4mm}
\item[Figure 8]
Spectroscopic properties of the $1\,^3\Sigma_g^+$ state of $^{7,7}$Li$_2$.
\end{description}
\end{center}
\end{figure}


\begin{thebibliography}{10}
\bibitem{dataPAS}
Three PAS data, those for $\,v(^{7,7}$Li$_2)= 62,~N=1)$, $\,v(^{6,6}$Li$_2)=
80,~N=1)\,$ and $\,v(^{6,6}$Li$_2)= 68,~N=2)\,$ proved to be outliers, and
were omitted from the final fits.
\end{thebibliography}
\end{document}